\documentclass[twocolumn,tighten]{aastex631}
\usepackage{graphicx}
\usepackage{amsmath}
\usepackage{natbib}
\usepackage{amssymb}
\shorttitle{Neutrino-Driven Winds from Magnetized Protoneutron Stars}
\shortauthors{D.~K.~Desai, D.~M.~Siegel, \& B.~D.~Metzger}

\begin{document}

\newcommand{\be}{\begin{equation}}
\newcommand{\ee}{\end{equation}}

\newcommand  \gcc {~\mathrm{g}~\mathrm{cm}^{-3}}
\newcommand  \E {\times 10^}

\title{Three-Dimensional General-Relativistic Simulations of Neutrino-Driven Winds from Magnetized Proto-Neutron Stars}

\author[0000-0002-8914-4259]{Dhruv K.~Desai}
\affil{Department of Physics and Columbia Astrophysics Laboratory, Columbia University, Pupin Hall, New York, NY 10027, USA}
\author[0000-0001-6374-6465]{Daniel M.~Siegel}
\affiliation{Institute of Physics, University of Greifswald, D-17489 Greifswald, Germany}
\affiliation{Perimeter Institute for Theoretical Physics, Waterloo, Ontario N2L 2Y5, Canada}
\affiliation{Department of Physics, University of Guelph, Guelph, Ontario N1G 2W1, Canada}
\author[0000-0002-4670-7509]{Brian D.~Metzger}
\affil{Department of Physics and Columbia Astrophysics Laboratory, Columbia University, Pupin Hall, New York, NY 10027, USA}
\affil{Center for Computational Astrophysics, Flatiron Institute, 162 5th Ave, New York, NY 10010, USA} 

\begin{abstract} 
Formed in the aftermath of a core-collapse supernova or neutron star merger, a hot proto-neutron star (PNS) launches an outflow driven by neutrino heating lasting for up to tens of seconds.  Though such winds are considered potential sites for the nucleosynthesis of heavy elements via the rapid neutron capture process ($r$-process), previous work has shown that unmagnetized PNS winds fail to achieve the necessary combination of high entropy and/or short dynamical timescale in the seed nucleus formation region.  We present three-dimensional general-relativistic magnetohydrodynamical (GRMHD) simulations of PNS winds which include the effects of a dynamically strong ($B \gtrsim 10^{15}$ G) dipole magnetic field.  After initializing the magnetic field, the wind quickly develops a helmet-streamer configuration, characterized by outflows along open polar magnetic field lines and a ``closed'' zone of trapped plasma at lower latitudes.  Neutrino heating within the closed zone causes the thermal pressure of the trapped material to rise in time compared to the polar outflow regions, ultimately leading to the expulsion of this matter from the closed zone on a timescale of $\sim$60 ms, consistent with the predictions of \citet{Thompson03}.  The high entropies of these transient ejecta are still growing at the end of our simulations and are sufficient to enable a successful 2nd-peak $r$-process in at least a modest $\gtrsim 1\%$ of the equatorial wind ejecta.    
\end{abstract}


\section{Introduction}
\label{sec:intro}

Magnetars are neutron stars with exceptionally high surface magnetic field strengths $\sim 10^{14}-10^{15}$ G (e.g., \citealt{kouveliotou_x-ray_1998,kaspi_magnetars_2017}), which comprise $\gtrsim 30\%$ of the young neutron star population \citep{Woods&Thompson06,beniamini_magnetar_evo_2019}.  These strong magnetic fields affect the appearance of magnetars throughout their active lifetimes, for example by providing an additional source of heating due to magnetic dissipation and by powering transient outbursts and flares (e.g., \citealt{zelati_outburst_2018, beniamini_magnetar_evo_2019,chime_burst_2020}).  As we shall explore in this paper, such strong magnetic fields can also affect the very first moments in a magnetar's life (e.g., \citealt{Thompson03}).

Whether formed from the core-collapse of a massive star, the accretion-induced collapse of a white dwarf (e.g., \citealt{dessart_multi-dimensional_2006}), or the merger of two neutron stars (e.g., \citealt{dessart_neutrino_2009,giacomazzo_formation_2013,perego_neutrino-driven_2014,Kaplan+14,Metzger+18}), all neutron stars begin as hot ``proto-neutron stars'' (PNS; e.g., \citealt{burrows_birth_1986,Pons+99}), which cool and contract via optically-thick neutrino emission for the first tens of seconds following their creation.  As neutrinos stream outwards from the neutrinosphere through the atmosphere of the PNS, the heat they deposit in the surface layers drives a thermally-driven outflow of baryons, known as the ``neutrino-driven'' wind (e.g., \citealt{Duncan+86,Qian&Woosley96,Thompson+01}).  
Neutrino-driven winds have received extensive interest as potential sites for the nucleosynthesis of rare heavy isotopes, particularly via the rapid neutron capture process ($r$-process; e.g., \citealt{Meyer+92}; \citealt{takahashi_nucleosynthesis_1994}; \citealt{woosley_r-process_1994}).  However, previous studies have shown that the roughly spherically symmetric winds from (slowly-rotating, unmagnetized) PNS fail to achieve the requisite combination of high entropy and short outflow expansion time through the seed nucleus formation region (\citealt{Hoffman+97,Meyer&Brown97,Meyer+92}) to enable the high ratio of neutrons to seeds necessary to achieve a successful 2nd or 3rd peak $r$-process (e.g., \citealt{Kajino+00,Sumiyoshi+00,Otsuki+00,Thompson+01,Arcones+07,fischer_protoneutron_2010,Roberts+10,Arcones&Montes11,Roberts+12a,MartinezPinedo+12,Fischer+12}).  Waves driven by convection, which steepen above the PNS surface and deposit additional entropy in the outflow, offer one possible mechanism to boost their $r$-process potential (e.g., \citealt{Suzuki&Nagataki05,metzger_proto-neutron_2007,Gossan+20,Nevins&Roberts23}).

As prefaced above, most previous studies of PNS winds also neglect the impact of two other neutron star properties: rotation and magnetic fields.  In recent work (\citealt{desai_pnsw_2022} – hereafter Paper I) we explored the effects of rapid rotation on the wind properties by means of three-dimensional general-relativistic (GR) hydrodynamic simulations.  These simulations revealed that, while rapid rotation (spin periods $P \lesssim$ few ms) acts to increase the mass-loss rate of the PNS near the rotational equator, the entropy and velocity of such outflows are suppressed compared to an otherwise equivalent non-rotating star, precluding $r$-process production via an alpha-rich freeze-out.  On the other hand, extremely rapid rotation approaching centrifugal break-out ($P \sim 1$ ms), was found to reduce the outflow electron fraction compared to the equivalent non-rotating wind model, thus potentially enabling the production of light $r$-process or light elementary primary process (LEPP) nuclei even absent an alpha-rich freeze-out.  

The prevalence of such very rapidly spinning PNS in nature is not clear, however.  Helioseismology observations and theoretical modeling suggest that angular momentum may be transferred out of the core of a massive star prior to core collapse with higher efficiency than had been previously assumed (e.g., \citealt{Cantiello_asteroseismic_2016,Fuller+19}).  Another potential consequence of rapid birth rotation, which may limit its prevalence in nature, is to endow the PNS with a strong magnetic field $\gtrsim 10^{15}$ G via a dynamo process (e.g., \citealt{Thompson&Duncan93,Price&Rosswog06,siegel_magnetorotational_2013, mosta_dynamo_2015, Margalit+22,White+22}).  While such ``millisecond magnetars'' are a contender for the central engines behind energetic transients such as engine-powered supernovae and gamma-ray bursts (e.g., \citealt{usov_millisecond_1992-1,wang_bipolar_sne_2001,Thompson+04,bucciantini_magnetar-driven_2007,metzger_proto-neutron_2007,bucciantini_magnetized_2009,Metzger+11a,Prasanna+23}), these explosions are so rare as to represent at most only a tiny fraction of all magnetar births (e.g., \citealt{Vink&Kuiper06,Fuller&Lu22}).  

On the other hand, rapid rotation may not be the only mechanism to generate magnetar-strength fields, which in principle could originate from flux freezing from the pre-collapse stellar core \citep{woltjer_1964,ruderman_1972, ferrario_fossil_2006,  Cantiello_asteroseismic_2016, white_origin_bfield_2022} or dynamos during the PNS phase which require less extreme core angular momentum \citep{barrere_2022}.  This motivates considering the effects of a strong magnetic field on PNS birth even absent rapid rotation. 

In this work (Paper II) we perform general-relativistic magnetohydrodynamical (GRMHD) simulations which explore the impact of strong ($B \gtrsim 10^{14}-10^{15}$ G) dipole magnetic fields on the properties of neutrino-heated outflows from non-rotating PNS winds.  We are motivated by the work of \citet{Thompson03}, who showed that a magnetar-strength field can initially trap plasma close to the PNS surface, where it can be heated by neutrinos to a higher entropy than achieved in an otherwise equivalent freely outflowing wind.  \citet{Thompson&udDoula18} studied this problem numerically by means of 2D axisymmetric MHD simulations with Newtonian gravity, using an approximate inner boundary condition near the PNS neutrinosphere and a prescribed free-streaming neutrino radiation field above.  They found that a moderate fraction $\sim 1-10\%$ of the wind material may be ejected in high-entropy outflows capable of $r$-process via an alpha-rich freezeout.  Likewise, \citet{Prasanna+22} performed similar simulations but covering a wider parameter space of the neutrino-driven winds from slowly rotating magnetars, focusing on the effects that neutrino-driven mass-loss in opening magnetic field lines and enhancing the star's spin-down rate relative to the magnetic dipole rate.  Our goal here is to explore a similar setup using 3D GRMHD simulations including neutrino transport via an M0 scheme, thus paving the way for future simulations within this numerical setup that include both magnetic fields and rapid rotation together.

This paper is organized as follows.  In Section \ref{sec:methods} we introduce the numerical code and models run. In Section \ref{sec:results}, we compare outflows from magnetars of various magnetic field strengths. In Section \ref{sec:conclusions} we summarize and compare our results to outflows from un-magnetized winds, and further discuss viability of $r$-process nucleosynthesis in magnetar winds.

\section{Methodology}
\label{sec:methods}

\subsection{Numerical Evolution Code, PNS Initial Data \& Grid Setup}
\label{sec:numerical}

We perform GRMHD simulations of magnetized PNS winds using a modified version of \texttt{GRHydro} \citep{mosta_grhydro_2014} as described in \citet{Siegel&Metzger18}, built on the open-source \texttt{Einstein Toolkit}\footnote{\href{http://einsteintoolkit.org}{http://einsteintoolkit.org}}\citep{goodale_cactus_2003,schnetter_evolutions_2004,thornburg_black-hole_2004,Loffler+12,babiuc-hamilton_einstein_toolkit_2019}. As in Paper I, we do not evolve spacetime for the current models. Instead, we evolve the GRMHD equations on a fixed background metric, determined at the start of the simulation by our initial data solver, for computational efficiency.  This assumption is justified by the fact that only a tiny fraction $\sim 10^{-5}$ of the star's mass is removed over the course of the simulation and the total energy in the magnetic field is $\lesssim 10^{-7}$ of the star's gravitational energy. 

We include weak interactions via a leakage scheme based on the formalism of \citet{bruenn_stellar_1985} and \citet{ruffert_coalescing_1996}, following the implementation in \citet{galeazzi_implementation_2013} and \citet{radice_dynamical_2016} (see also \citealt{Siegel&Metzger18}). In the presence of strong magnetic fields, the neutrino heating and cooling rates will be altered with respect to a non-magnetized setup, due to the impact of Landau quantization on the available electron/positron states (e.g., \citealt{lai_neutrino_1998,duan_neutrino_2004}). However, as we show in Appendix \ref{appendix:landau}, these effects are small, even for the strongest magnetic field strength cases explored in this work.  We are thus justified in neglecting these corrections.

As in Paper I, we employ a one-moment approximation to the general-relativistic Boltzmann equation (`M0') to describe neutrino transport. It is implemented as a ray-by-ray scheme \citep{radice_dynamical_2016} in our enhanced version of \texttt{GRHydro}, in which neutrino mean energies and number densities are evolved along null radial coordinate rays (see also \citealt{Combi2022}). The M0 radiation transport grid is separate from that of the GRMHD variables. It uses spherical coordinates centered on the PNS, uniformly spaced in $\{r,\theta,\phi\}$ extending out to radii of 200\,km, with the number of grid points $(n_r,n_\theta,n_\phi)=(600,20,40)$. It thus covers the density range $\rho \gtrsim 10^4 \gcc$, for which weak interactions are relevant, to ensure accuracy of the model (see Paper I for more details). 

%The initial conditions for the PNS structure are determined by the TOV equations, with a realistic, density-dependent initial temperature and $Y_e$ profile based on \citet{Kaplan+14}, as implemented in Paper I.  We use the tabulated SFHo equation of state \citep{oconnor_new_2010}\footnote{Available in tabulated form on \hyperlink{stellarcollapse.org}{stellarcollapse.org}.} introduced by \citet{Steiner+13}. 

Our magnetized PNS models use as initial conditions unmagnetized PNS wind solutions similar to those presented in Paper I, run sufficiently long for the wind to achieve a quasi-steady-state, but short compared to the Kelvin-Helmholtz cooling timescale of the PNS. These unmagnetized wind solutions are then endowed with a large-scale dipolar magnetic field (see Sec.~\ref{sec:Bfield_setup} for more details). Here, we briefly review the setup of the unmagnetized models. 

The initial conditions for the PNS structure are determined by the Tolman-Oppenheimer-Volkoff equations, employing a density-dependent initial temperature and $Y_e$ profile (computed by imposing $\beta$-equilibrium), following the approach by \citet{Kaplan+14}. We use the SFHo equation of state (\citealt{Steiner+13}) in tabular form as provided by \citet{oconnor_new_2010}\footnote{\href{https://stellarcollapse.org/equationofstate.html}{https://stellarcollapse.org/equationofstate.html}}. We do not self-consistently follow PNS cooling evolution to arrive at initial conditions for our models; thus the final electron fraction $Y_e$ in the outflows, which depends sensitively on the neutrinosphere properties, will not be physically accurate.  Nevertheless, we can draw conclusions regarding the effects (if any) of strong magnetic fields on the wind composition by comparing $Y_e$ from our magnetized models to those obtained from otherwise equivalent unmagnetized models.

Our fiducial model employs a grid hierarchy consisting of a base Cartesian grid and six fixed nested refinement levels. We improve upon the resolution used in the models from Paper I. The smallest and finest grid is a $15 \times 15 \times 15$\,km box centered at the origin now with a resolution of 150\,m, while the size of the base grid is 960 $\times$ 960 $\times$ 960\,km. As discussed in Paper I, this grid setup allows us to simultaneously capture the wind zone on scales of ${\sim}100-1000$\,km, while providing enough resolution from the finest grid to capture the details of outflows from the PNS surface. %We find this level of resolution necessary to evolve the magnetic field in a stable manner. For lower resolutions, we find violations of the $\nabla \cdot \vec{B} = 0$ contraint, as well as errors from regions with low $\beta \equiv P_f/P_B$ \citep{siegel_recovery_2018}, where $P_f$ and $P_B$ are the fluid and magnetic pressure, respectively (fluid pressure is determined from the equation of state). 
For this resolution, the magnetic pressure scale height $H_{p_B}=|d \ln p_B/dr|^{-1}$ is resolved by at least 10 grid points throughout the entire domain of the simulation
%, and by at least 10 grid points at and exterior to the neutrinosphere
(see Appendix~\ref{appendix:resolution}, Fig.~\ref{fig:mag_scale_height}). As in Paper I, the neutrinosphere is only marginally resolved, with the grid resolution approximately equal to the optical depth scale height (see Appendix~\ref{appendix:resolution}, Fig.~\ref{fig:mag_scale_height}). This issue is not critical for the purposes of this study, as the neutrino energies and luminosities determined at the neutrinosphere serve as a boundary condition for heating that occurs at larger radii, a region that by comparison is well resolved. Most of our calculations employ reflection symmetry across the equatorial ($z = 0$) plane for computational efficiency.  We have checked that our results for the time-averaged wind properties are not appreciably affected by this assumption, by performing a full-domain simulation and a side-by-side comparison to the half-domain case (see Appendix~\ref{appendix:symmetry}).

\subsection{Addition of a Dipole Magnetic Field}
\label{sec:Bfield_setup}

\begin{table*}[ht!]
\begin{center}
    \begin{tabular}{|c|c|c|c|c|c|c|c|}
    \hline
     Model & $B_{\rm S}^{(a)}$ & $R_e^{(b)}$  & $R_{\bar \nu_e}^{(c)}$ & $\langle L_{\nu_e} \rangle^{(d)}$ & $\langle L_{\bar{\nu}_e} \rangle^{(e)}$ & $\langle E_{\nu_e}\rangle^{(f)}$ & $\langle  E_{\bar{\nu}_e} \rangle^{(g)}$ \\
- &($10^{15}$ G) & (km) & (km) & (erg s$^{-1}$) & (erg s$^{-1}$) & (MeV) & (MeV)\\
      \hline \hline

\texttt{no-B}$^*$ & 0.0 & 13 & 10 & 2.2e51 & 3.9e51 & 13 & 18\\

\texttt{lo-B} & 0.61 & 13  & 10 & 2.2e51 & 3.9e51 & 13 & 18\\

\texttt{hi-B} & 2.5 & 13  & 10 & 2.3e51 & 3.8e51 & 13 & 18\\
\hline

\texttt{sym-B}$^{\dagger}$ & 2.1 & 13  & 11 & 4.3e51 & 6.0e51 & 13 & 17\\
\texttt{no-sym-B}$^{\dagger}$ & 2.2 & 13  & 11 & 4.1e51 & 6.8e51 & 13 & 18\\

\hline
    \end{tabular}\\
    \end{center}
    \vspace{-4mm}
    \caption{\textbf{Suite of PNS Wind Simulations: Magnetic Field and Neutrino Properties } \label{tab:models} \\
    ${(a)}$ Magnetic field strength as measured along the polar axis at the radius of the $\bar{\nu}_e$ neutrinosphere;
    ${(b)}$ Initial equatorial radius of the star; 
    ${(c)}$ Steady-state $\bar{\nu}_e$ neutrinosphere radius;
    ${(d)-(g)}$ Luminosities and mean energies of electron neutrinos and anti-neutrinos, averaged over the final factor of three in simulation time.\\
    $^*$Fiducial unmagnetized wind model (Fig.~\ref{fig:no-B}, left panels).
    $^{\dagger}$All simulations employ the same grid geometry, but these models are run with three times poorer resolution compared to the models in the first 3 rows (i.e., $\Delta x = 450$ m for the finest grid).}
\end{table*}

\begin{figure}
\includegraphics[width=0.48\textwidth]{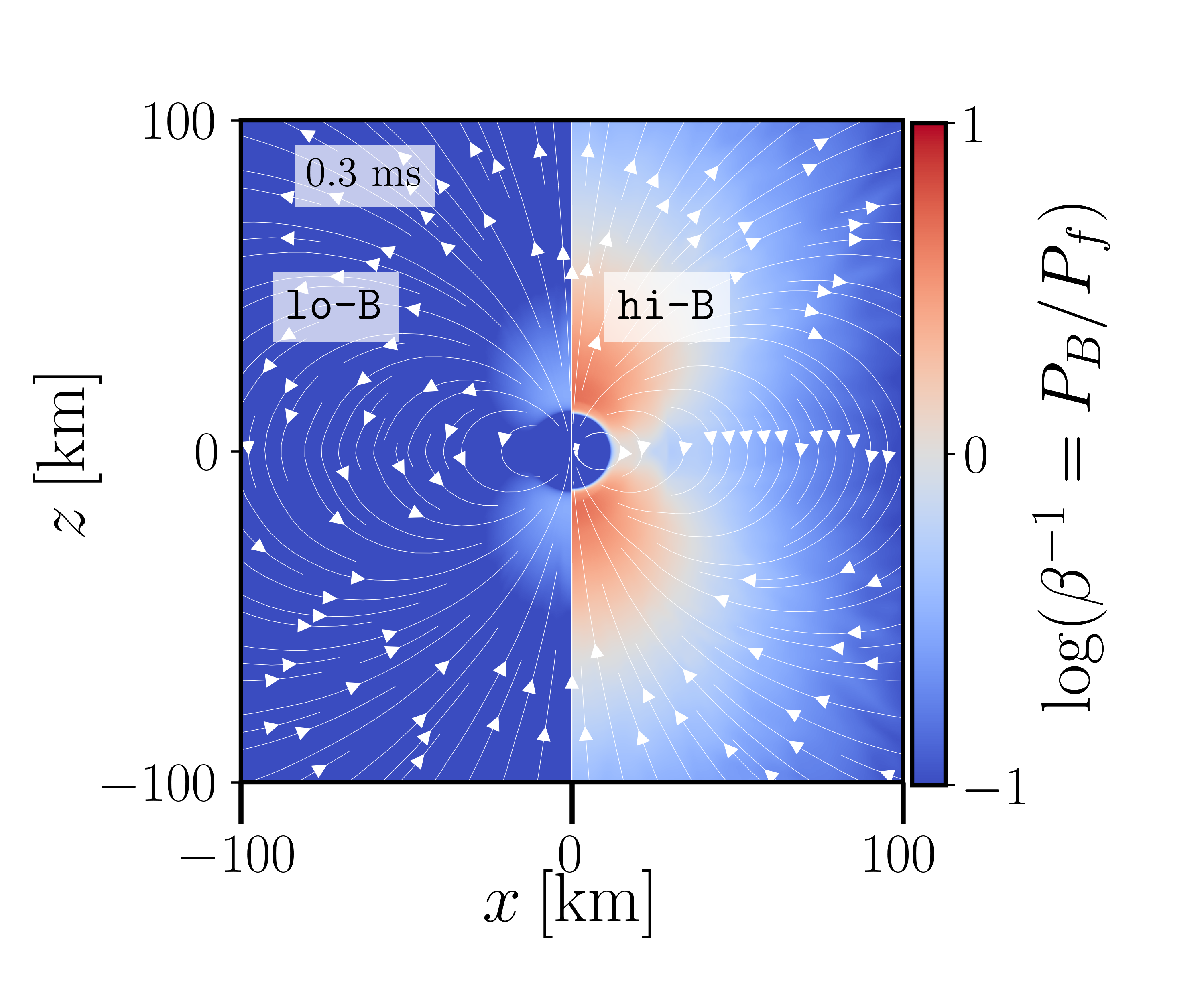}
\caption{Initial magnetic-to-fluid pressure ratio $\beta^{-1}$ with magnetic field line geometry overlaid.  We show models \texttt{lo-B} (left) and \texttt{hi-B} (right) in a slice through the magnetic dipole axis ($y=0$ plane), soon after the magnetic field is turned on at $t=0$ ms. This magnetic field configuration was added onto an unmagnetized PNS wind solution previously evolved for about $176$\,ms to steady state. For model \texttt{hi-B}, magnetic pressure dominates over fluid pressure in the polar regions above the PNS surface out to $\approx \!60$\,km.
}
\label{fig:initial_mhd}
\end{figure}

We evolve the unmagnetized PNS and its wind for ${\sim}150$\,ms, the last ${\sim}50$\,ms over which the wind has achieved a quasi-steady state. At this point, we superimpose a large scale magnetic field onto the stationary wind solution. As in \citet{siegel_magnetically_2014} and similar to the configurations in \citet{shibata_afterglow_2011} and \citet{kiuchi_three-dimensional_2012}, we define this large-scale, dipole-like magnetic field by initializing the components of the vector potential as $A_r=A_\theta = 0$,
\be
A_\phi = A_{0,d} \frac{\varpi}{(r^2 + \varpi^2_{0,d}/2)^{3/2}},
\ee 
where $A_{0,d}$ tunes the overall field strength, $\varpi$ is the cylindrical radius, $r^2 = \varpi^2 + z^2$, and $\varpi_{0,d} \simeq 8$ km is the `neutral point'. The neutral point corresponds to a ring-like current in the equatorial ($z=0$) plane centered at the origin. We require this neutral point to lie within the star (of radius $R_{\rm PNS} \simeq 12$ km) so as to yield a plausible, nearly point-like dipole field that is non-singular everywhere. The magnetic field is then obtained via $B=\nabla \times A $, yielding the initial configuration shown in Fig.~\ref{fig:initial_mhd}. The dipole geometry is symmetric across the $z=0$ plane at initialization. Although the current sheet that forms in this plane can ``wobble", we find that when averaged over time, these wobbles smooth out (Fig.~\ref{fig:ref_sym} of Appendix~\ref{appendix:symmetry}).  Hence, the time-averaged wind outflow properties should also be roughly symmetric across the $z=0$ plane, justifying our use of equatorial symmetry for computational efficiency.

We activate the magnetic field after running the non-magnetized PNS wind evolution for roughly $\approx 176$\,ms, at which point we ``reset'' the clock and hereafter refer to this time as $t=0$.  In order to explore different physical regimes of magnetized winds, we consider models which span two different initial surface magnetic field strengths but are otherwise identical.  We distinguish these models by their polar magnetic field strength $B_{\rm S}$ at the neutrinosphere, with $B_{\rm S} \simeq [6.1\E{14}$\,G, 2.5$\E{15}$\,G], referred to as \texttt{lo-B} and \texttt{hi-B}, respectively (see Tab.~\ref{tab:models} and Fig.~\ref{fig:initial_mhd} for further details). In addition to the magnetized models, we run in parallel a non-magnetized model ($B_{\rm S} = 0$, referred to as \texttt{no-B}), which provides a set of comparison wind properties over the same time interval (i.e., for the same PNS neutrino cooling evolution).

The magnetic field does not significantly impact the hydrostatic structure of the PNS.  At the time the field is initialized, the magnetic-to-fluid pressure ratio $\beta^{-1} \equiv P_B/P_f$ obeys $ < 10^{-2}$ for all the models we consider (see Fig.~\ref{fig:initial_mhd}).  This ratio similarly shows that the gradient of the magnetic field is negligible with respect to the fluid pressure gradient.  The magnetic field is not dynamically relevant; only the wind region is potentially impacted by its presence.  

The toroidal component of the magnetic field remains subdominant compared to the poloidal component at all epochs and for all models (see Appendix~\ref{appendix:toroidal}). This is expected because the initial magnetic field configuration is purely poloidal and there is no significant active mechanism, such as rotation, winding up magnetic fields on large scales.

\begin{figure}
%\centering
\includegraphics[width=.48\textwidth]{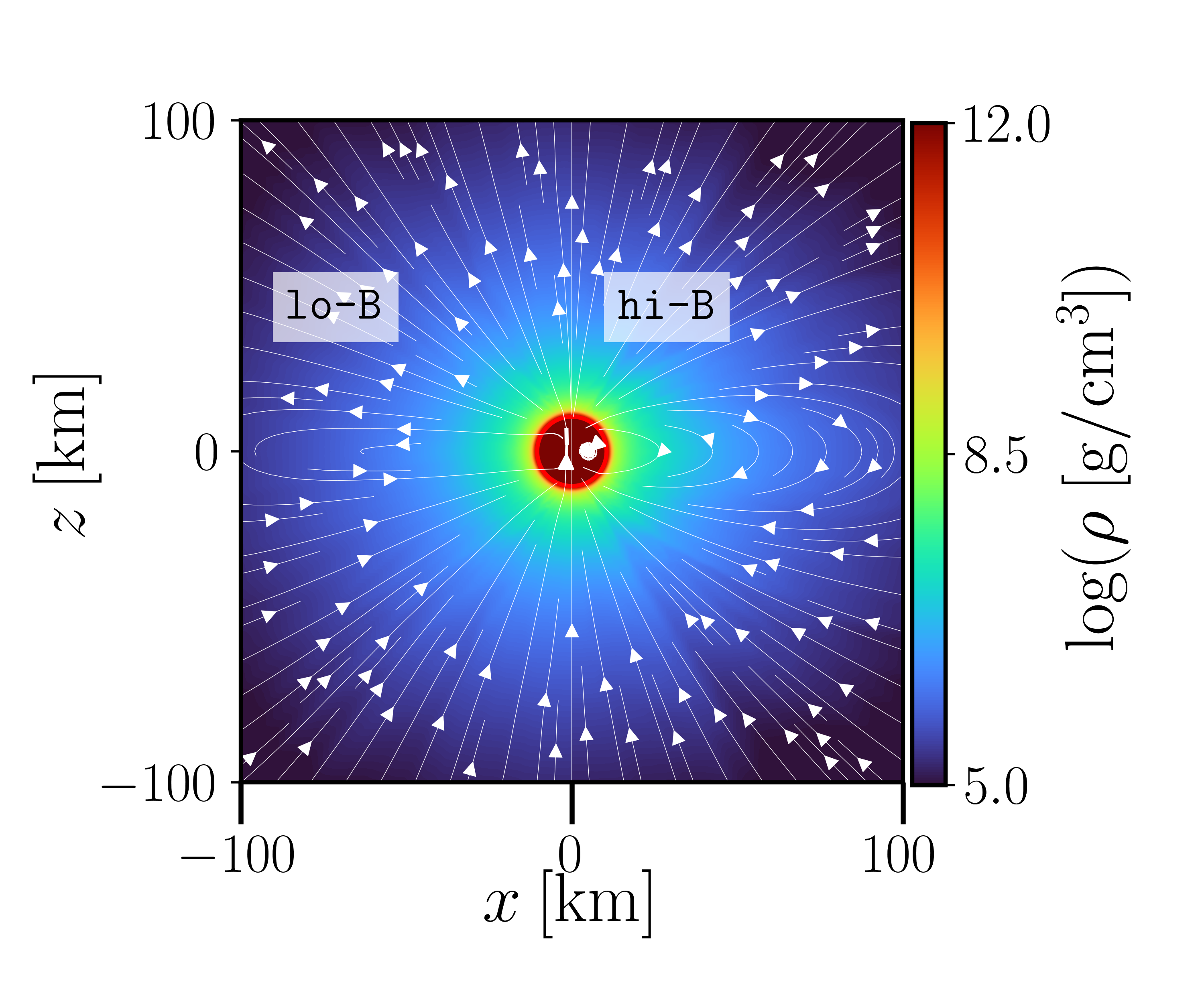}
\caption{Snapshots of the outflow properties as a cross-section through the magnetic dipole axis ($y = 0$ plane) at $t \approx 55$\,ms after $B$-field initialization for models \texttt{lo-B} (left) and \texttt{hi-B} (right).  Colors show the rest-mass density $\rho$, with the magnetic field line geometry overlaid in white. The neutrinosphere ($\tau_{\bar \nu_e}=1$ surface) is shown with a red contour.  The field lines have been nearly completely torn open radially in the weakly magnetized model, while an equatorial belt of closed field lines persists in the strongly magnetized case.}
\label{fig:split_mono}
\end{figure}

\begin{figure}
%\centering
\includegraphics[width=.48\textwidth]{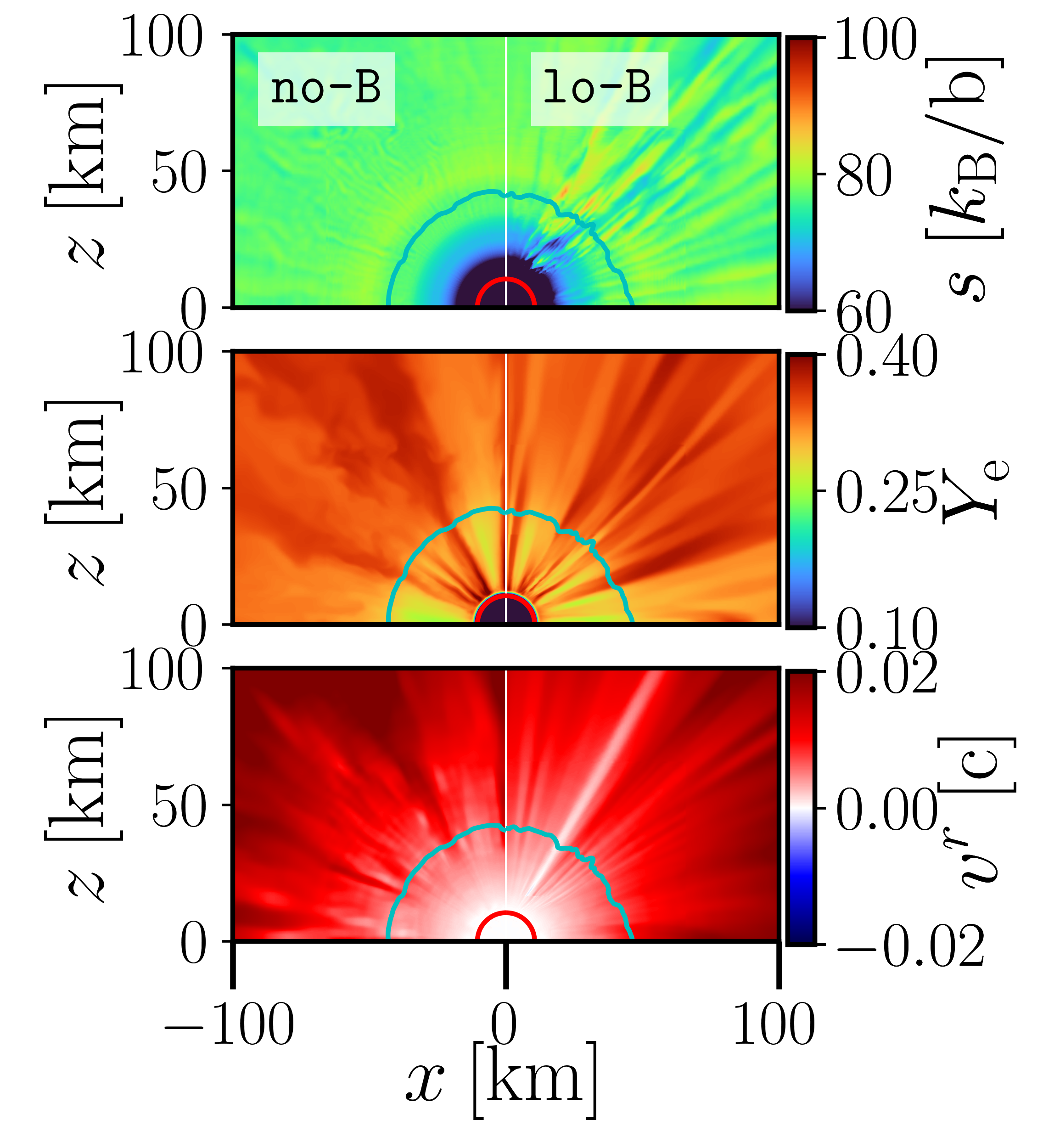}
\caption{Snapshots showing outflow properties as a cross-section through the magnetic dipole axis ($y = 0$ plane) at $t \approx 55$\,ms after $B$-field initialization for models \texttt{no-B} (unmagnetized, left) and \texttt{lo-B} (weakly magnetized, right). From top to bottom: specific entropy $s$, electron fraction $Y_e$, radial velocity $v^r$. The green contour represents the alpha-particle formation surface ($X_\alpha=0.5$), around which $r$-process seed nuclei begin to form, while the red contour represents the location of the neutrinosphere ($\tau_{\bar \nu_e}=1$ surface).}
\label{fig:no-B}
\end{figure}

\section{Results}
\label{sec:results}

The presence of a magnetic field can significantly impact the dynamics of the PNS wind in regions where the magnetic pressure $P_{\rm B} = B^{2}/8\pi$ greatly exceeds the fluid pressure $P_f$ (\citealt{thompson_magnetic_2003,Thompson&udDoula18,Prasanna+22}).  In the gain region just above the PNS surface, where neutrino heating begins to exceed neutrino cooling, radiation pressure of photons and electron/positrons dominates over gas pressure and hence $P_f \simeq P_{\rm rad} = (11/12)aT^4$, where $a$ is the radiation constant.  Just below this point closer to the star, the specific neutrino heating rate $\propto L_{\nu}\epsilon_{\nu}^{2}/r^{2}$ and neutrino cooling rate $\propto T^{6}$ (due to pair capture reactions on free nucleons) balance, resulting in a roughly isothermal atmosphere of temperature (e.g., \citealt{qian_nucleosynthesis_1996})
\be
T_{\rm eq} \approx 1~L_{\bar \nu_e, 51}^{1/6}\langle E_{\bar \nu_e, \rm MeV}\rangle^{1/3}R_{\bar \nu_e,6}^{-1/3} \rm~MeV , 
\label{eq:temp_eq}
\ee
where $R_{\bar \nu_e} = R_{\bar \nu_e,6}\times 10^{6}$ cm , $L_{\bar \nu_e} = L_{\bar \nu_e,51}\times 10^{51}$ erg s$^{-1}$, and $\langle E_{\bar \nu_e}\rangle = \langle E_{\bar \nu_e, \rm MeV}\rangle \times $ MeV are the PNS radius, electron antineutrino luminosity, and mean electron antineutrino energy, respectively.

Equation~\ref{eq:temp_eq} reveals that $P_{\rm B} > P_{\rm rad}$ for magnetic field strengths above a critical value
\be
B_{\rm crit} \approx 7\E{14}\,{\rm G}\mskip-3mu\left( \frac{L_{\bar \nu_e,51}}{4} \right)^{\frac{1}{3}}\mskip-5mu
\left(\frac{\epsilon_{\bar \nu_e}}{18\,\mathrm{MeV}}\right)^{\frac{2}{3}}
\mskip-5mu\left(\frac{R_{\bar \nu_e}}{10\,\mathrm{km}} \right)^{-\frac{2}{3}}\mskip-10mu .
\label{eq:Bcrit}
\ee
Thus, for surface dipole fields $B_{\rm S} \gg B_{\rm crit}$ we expect the wind dynamics to be significantly altered by the presence of the magnetic field.

%As mentioned in Sec.~\ref{sec:numerical}, sufficiently strong magnetic fields can impact the weak interaction rates by quantization of the electron/positrons phase space into Landau levels \citep{lai_neutrino_1998,Duan&Qian04,duan_rates_2005}.  These effects become important if $\hbar \Omega_B > p_e$, where $\Omega_B$ is the cyclotron frequency, $\hbar$ is Planck's constant, and $p_e$ is the Fermi momentum.  Given gain layer temperatures $\approx 1$ MeV (Eq.~\ref{eq:temp_eq}), this requires a magnetic field strength of X, much stronger than typical magnetic fields. We show in Fig.~\ref{fig:landau} of the Appendix that for all densities relevant to the simulations in this study, temperatures remain below the critical temperature for Landau effects to . Thus we neglect such effects of the B-field on the microphysics of the wind in this study.

We begin in Sec.~\ref{sec:weakmag} by presenting results for a relatively weakly magnetized model (\texttt{lo-B}), for which $B_{\rm S} < B_{\rm crit}$ and hence the wind dynamics are expected to be similar to the unmagnetized wind. In Sec.~\ref{sec:strongmag}, we move onto describing a strongly magnetized model (\texttt{hi-B}) for which $B_{\rm S} > B_{\rm crit}$ and discuss prospects for $r$-process nucleosynthesis in Sec.~\ref{sec:nucleosynthesis}.

\subsection{Weakly Magnetized Model}
\label{sec:weakmag}
The left panel of Fig.~\ref{fig:initial_mhd} shows the magnetic-to-fluid pressure ratio $\beta^{-1} = P_{\rm B}/P_{\rm f}$ and magnetic field line geometry for the weakly magnetized model \texttt{lo-B} at the moment of magnetic field initialization. Figure~\ref{fig:no-B} shows a comparison of key quantities from the magnetized and unmagnetized models (\texttt{lo-B} and \texttt{no-B}, respectively) 55\,ms after a weak magnetic field is turned on.

Before the magnetic field is activated, the PNS wind has achieved an approximately steady state, as detailed in Paper I. At the moment the magnetic field is initialized, magnetic field lines reflect a dipole geometry and $\beta^{-1} = P_{\rm B}/P_{\rm f}$ of the wind is $\lesssim 1$, indicating that fluid pressure dominates over magnetic pressure from the start (Fig.~\ref{fig:initial_mhd}). As the wind subsequently evolves, the dipole magnetic field structure is disrupted, as evidenced by the field line geometry at late times (Fig.~\ref{fig:split_mono}, left panel). The magnetic field is `frozen in' within the radial fluid flow (see the velocity field in Fig.~\ref{fig:no-B}, bottom panel). By $\sim 55$ ms after magnetic field initialization, outflows in models \texttt{lo-B} ($B_{\rm S}=6.1\E{14}$ G) and \texttt{no-B} ($B_{\rm S}=0$) are nearly indistinguishable (Fig.~\ref{fig:no-B}).

\begin{figure}
\centering
\includegraphics[width=.5\textwidth]{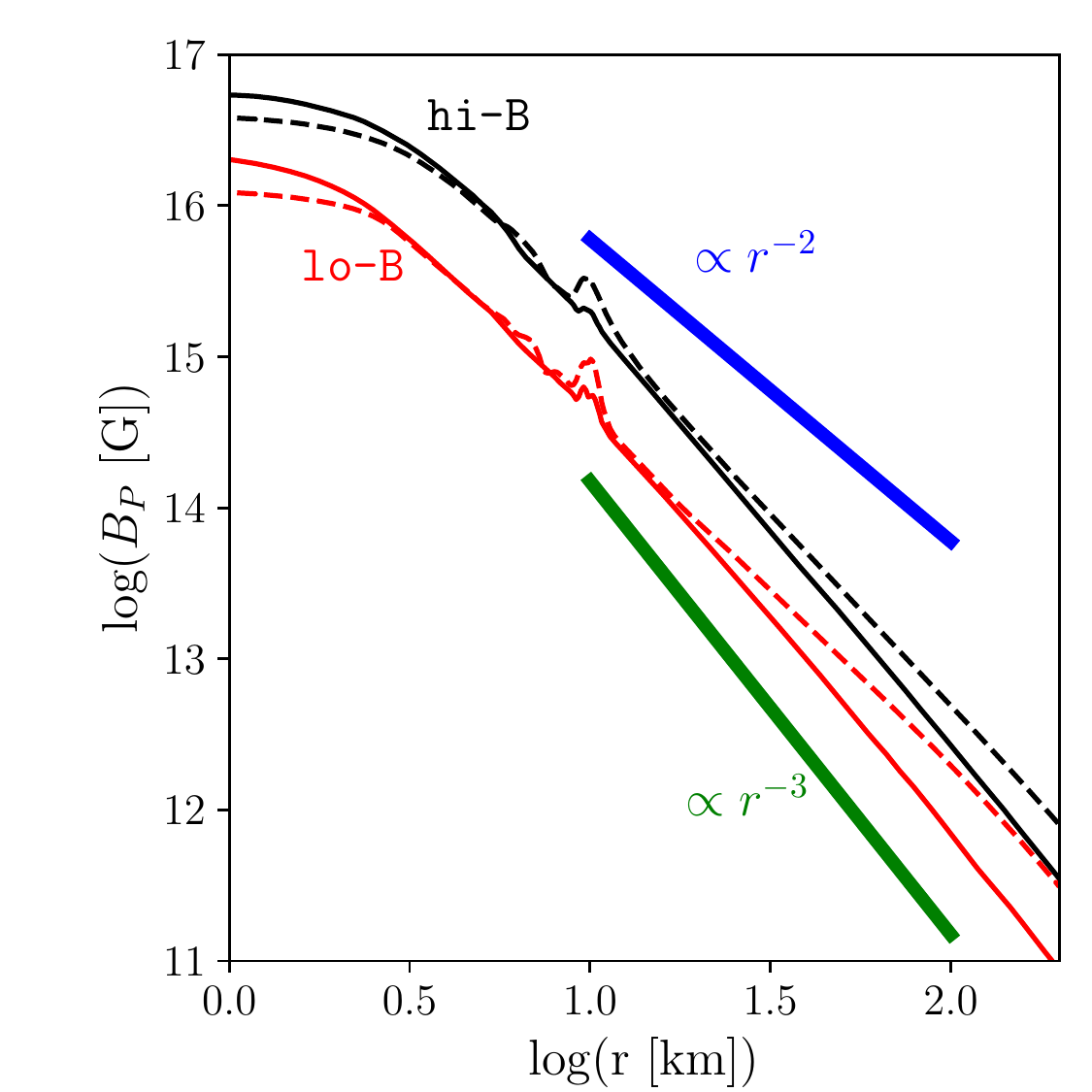}
\caption{Radial profiles of the poloidal magnetic field strength $B_{\rm P}$ for models \texttt{lo-B} and \texttt{hi-B}, taken from a slice through the magnetic dipole axis ($y=0$), angle-averaged in the polar region from $\theta=0^\circ$ to $\theta=45^\circ$.  Solid lines show $B_{\rm P}$ time-averaged from $t\approx 0-5$ ms after the magnetic field is activated, while dashed lines represent $B_{\rm P}$ time-averaged from $t\approx 25-35$ ms. To guide the eye, thick solid lines show power-laws $r^{-2}$ (blue) and $r^{-3}$ (green). Both models approach the split monopole solution ($\propto r^{-2}$) in the polar region.}
\label{fig:Bpol}
\end{figure}

Because fluid pressure exceeds magnetic pressure upon magnetic field initialization nearly everywhere ($\beta^{-1} <1$ in Fig.~\ref{fig:initial_mhd}, left), and since magnetic fields are frozen into the fluid flow, magnetic field lines follow the mostly radial outflow of the neutrino-driven wind. The magnetic field thus tears open from its initial dipole configuration within milliseconds, approaching a split monopole configuration (see left panel of Fig.~\ref{fig:split_mono}). For such a split-monopole solution, $B(r) \propto r^{-2}$, rather than $B(r) \propto r^{-3}$, as for the dipole geometry (e.g., \citealt{weber_davis_1967}). This is illustrated in Fig.~\ref{fig:Bpol}, which shows angle-averaged radial profiles of the poloidal magnetic field strength in the polar region for models \texttt{lo-B} and \texttt{hi-B} at the time the magnetic field is initialized (solid lines, for which $B(r) \propto r^{-3}$) as well as $\sim$30 ms later (dashed lines, for which $B(r) \propto r^{-2}$). A thin `current sheet' region forms in the equatorial plane with a low magnetic-to-fluid pressure ratio $\beta^{-1}$ as the wind evolves to $t\approx 55$ ms (Fig.~\ref{fig:split_mono}, left panel). Here, magnetic field lines of opposite polarities from the northern and southern hemispheres reconnect and heat the fluid (however, note that the physical thickness of the current sheet is not resolved by the simulation because the resistivity is numerical in this ideal MHD setup). 

Aside from the magnetic field properties, all of the asymptotic wind properties remain largely unaffected compared to the prior unmagnetized state. The PNS hydrostatic structure and neutrino energies as well as luminosities are essentially unchanged with respect to the unmagnetized model (see Tab.~\ref{tab:models}). Consequently, the net specific heating rate $\dot{q}_{\rm net}$, which depends on neutrino properties and temperature, retains a similar radial profile.  After an initial transient phase, the outflow properties of model \texttt{lo-B} such as the specific entropy $s$, electron fraction $Y_e$, and radial velocity $v^r$, and the asymptotic wind properties are also very similar to the \texttt{no-B} model (Fig.~\ref{fig:no-B}, Tab.~\ref{tab:wind_props}).

Because the temperature and density profiles change little after the magnetic field is activated, the surface at which $\alpha$-particles form (roughly where $T \lesssim 5 \times 10^9$ K) remains at $r\approx45$ km, close to the outer edge of the gain region (Fig.~\ref{fig:no-B}).  As outlined by \citet{qian_nucleosynthesis_1996} and reviewed in Paper I, the ability of the PNS wind to generate $r$-process nuclei yields depends on $s$, $Y_e$, and the expansion timescale $t_{\rm exp}$, which we define as
\be
t_{\rm exp} = \left(v^r\frac{\ln dT}{dr}\right)^{-1}\bigg|_{X_{\rm nuc}=0.5}, \label{eq:texp}
\ee
where we have defined $X_{\rm nuc}=0.5$ as the $\alpha$-particle surface ($X_{\rm nuc}$ is the mass fraction of all nuclei excluding protons and neutrons), as typically located $50-100$ km above the PNS surface. In particular, the $r$-process figure of merit defined by \citep{Hoffman+97}
\be
\eta \equiv \frac{s^3}{Y_e^3 t_{\rm exp}} \label{eq:eta}
\ee
is a rough measure of success for heavy element nucleosynthesis due to alpha-rich freeze-out, where here and hereafter $t_{\rm exp}$ is measured in seconds and $s$ is meausred in $k_{\rm B}$ per baryon.  Threshold values of $\eta \gtrsim 4\E{9}$ and $\eta \gtrsim 9\E{9}$ are required for neutron capture to proceed to the 2nd and 3rd $r$-process peak, respectively. As found in Paper I, and consistent with past findings (e.g., \citealt{qian_nucleosynthesis_1996,Hoffman+97,Thompson+01}), unmagnetized PNS winds (model \texttt{no-B}) generally have $\eta \ll 1\E{9}$, insufficient for an $r$-process.  Not surprisingly then, our weakly magnetized model \texttt{lo-B} is also not capable of an $r$-process (Tab.~\ref{tab:wind_props}).

\begin{table*}
%\centering
  \begin{center}
      %\caption{Time-Averaged Outflow Properties \label{tab:wind_props}}
    \begin{tabular}{|c|c|c|c|c|c|c|c|}
    \hline
     Model & Sector$^{\dagger \dagger}$ & $ \langle \dot M \rangle^{(a)}_{\rm 60~km}$ & $\langle \dot E_K \rangle_{\rm 60~km}^{(b)}$ & $\langle s\rangle^{(c)}_{\rm 60~km} \pm \sigma_s$
     & $\langle t_{\rm exp} \rangle^{(d)}_{\rm 60~km} \pm \sigma_{t_{\rm exp}}$ & $\langle Y_e\rangle^{(e)}_{\rm 60~km} \pm \sigma_{Y_e}$ & $\langle \eta \rangle^{(f)}_{\rm 60~km} \pm \sigma_{\eta}$\\
- & - & ($M_{\odot}$ s$^{-1}$) & (erg s$^{-1}$) & $k_B$ baryon$^{-1}$ & (ms) & - & $[10^8]$\\
      \hline \hline

\texttt{no-B}$^*$ & tot. & 5.5e-5 & 5.4e45 & 78$\pm$0.9 & 20$\pm$9 & 0.40$\pm$0.003 & 5.6$\pm$2 \\

\texttt{lo-B} & tot. & 5.6e-5  & 5.1e45 & 78$\pm$1.9, (78$\rightarrow$78) & 21$\pm$6 & 0.36$\pm$0.003 & 5.3$\pm$2 (5.2$\rightarrow$5.3) \\

\texttt{hi-B} & tot. & 3.8e-5  & 3.1e45 & 82$\pm$5 & 29$\pm$20 & 0.37$\pm$0.002 & 5.1$\pm$4 \\

\texttt{hi-B} & pol. & 2.9e-5 & 2.3e45 & 80$\pm$4 (80$\rightarrow$80) & 32$\pm$10 & 0.37$\pm$0.002 & 3.8$\pm$2 (3.5$\rightarrow$3.9) \\
\texttt{hi-B} & equat. & 4.3e-5  & 3.6e45 & 84$\pm$6 (83$\rightarrow$85) & 25$\pm$20 & 0.37$\pm$0.02 & 6.5$\pm$4 (5.7$\rightarrow$7.8) \\
\hline
\texttt{sym-B}$^{\dagger}$ & tot.  & 1.6e-4  & 3.6e46 &  67$\pm$3 & 16$\pm$10 & 0.38$\pm$0.02 & 4.6$\pm$2 \\
\texttt{no-sym-B}$^{\dagger}$ & tot. & 1.8e-4 & 5.4e46 & 68$\pm$3 & 21$\pm$40 & 0.38$\pm$0.03 & 4.6$\pm$3 \\

\hline
\end{tabular}
\end{center}
\vspace{-4mm}
    \caption{\textbf{Time-Averaged Outflow Properties} \label{tab:wind_props} \\
    We report wind properties time-averaged from $t\approx 25-55$ ms for models \texttt{no-B}, \texttt{lo-B}, and \texttt{hi-B}, and from $t\approx 0-20$ ms for models \texttt{sym-B} and \texttt{no-sym-B}. Entries of the format $(q\rightarrow q')$ refer to the same quantity time-averaged from $t\approx 25-35$ ms ($q$) and from $t\approx 45-55$ ms ($q'$).\\
    ${(a)}$ isotropic mass-loss rate; 
    ${(b)}$ kinetic energy;
    ${(c)}$ specific entropy with standard deviation;
    ${(d)}$ expansion timescale with standard deviation;
    ${(e)}$ electron fraction with standard deviation;
    ${(f)}$ $r-$process figure of merit $\eta \equiv s^3/(Y_e^3 t_{\rm exp})$ with standard deviation.\\
    $^*$Fiducial unmagnetized wind model shown in Fig.~\ref{fig:no-B} (left-hand panels).
    $^{\dagger}$All simulations in the table use the same grid geometry, but these models are run with a third of the resolution compared to the models in the first three rows (i.e.~$\Delta x = 450$ m for the finest grid).
    $^{\dagger \dagger}$Quantities are averaged over different angular sectors in polar angle $\theta$: `equat.' ($45^\circ \leq \theta \leq 135^\circ$), `pol.' ($0^\circ \leq \theta \leq 45^\circ$ and $135^\circ \leq \theta \leq 180^\circ$), and `tot.' ($0^\circ \leq \theta \leq 180^\circ$). Isotropic equivalent quantities are reported for $\langle \dot M \rangle$ and $\langle \dot E_K \rangle$ at a radius $r=60$ km.}
% \label{tab:wind_props}}
\end{table*}

\subsection{Strongly Magnetized Model}
\label{sec:strongmag}

We now consider results for the strongly magnetized model (\texttt{hi-B}) with $B_{\rm S}=2.5\E{15}\,{\rm G} > B_{\rm crit}$.  Immediately after the magnetic field is activated, its dipole geometry is identical to that of model \texttt{lo-B}, with the poloidal field strength following the expected $r^{-3}$ radial profile (Fig.~\ref{fig:Bpol}).  The magnetic-to-fluid pressure ratio, however, is much larger compared to the weakly magnetized model, reaching $\beta^{-1} \sim 10$ near the polar regions for model \texttt{hi-B} (Fig.~\ref{fig:initial_mhd}).  Right after the magnetic field is initialized, the radial velocity $v^r$ is negative in a region extending from slightly above the neutrinosphere at $r \approx 15$ km to $r \approx 50$ km, as a result of the previously outflowing wind plasma being trapped by the strong magnetic field.

\begin{figure}
\centering
    \includegraphics[width=.45\textwidth]{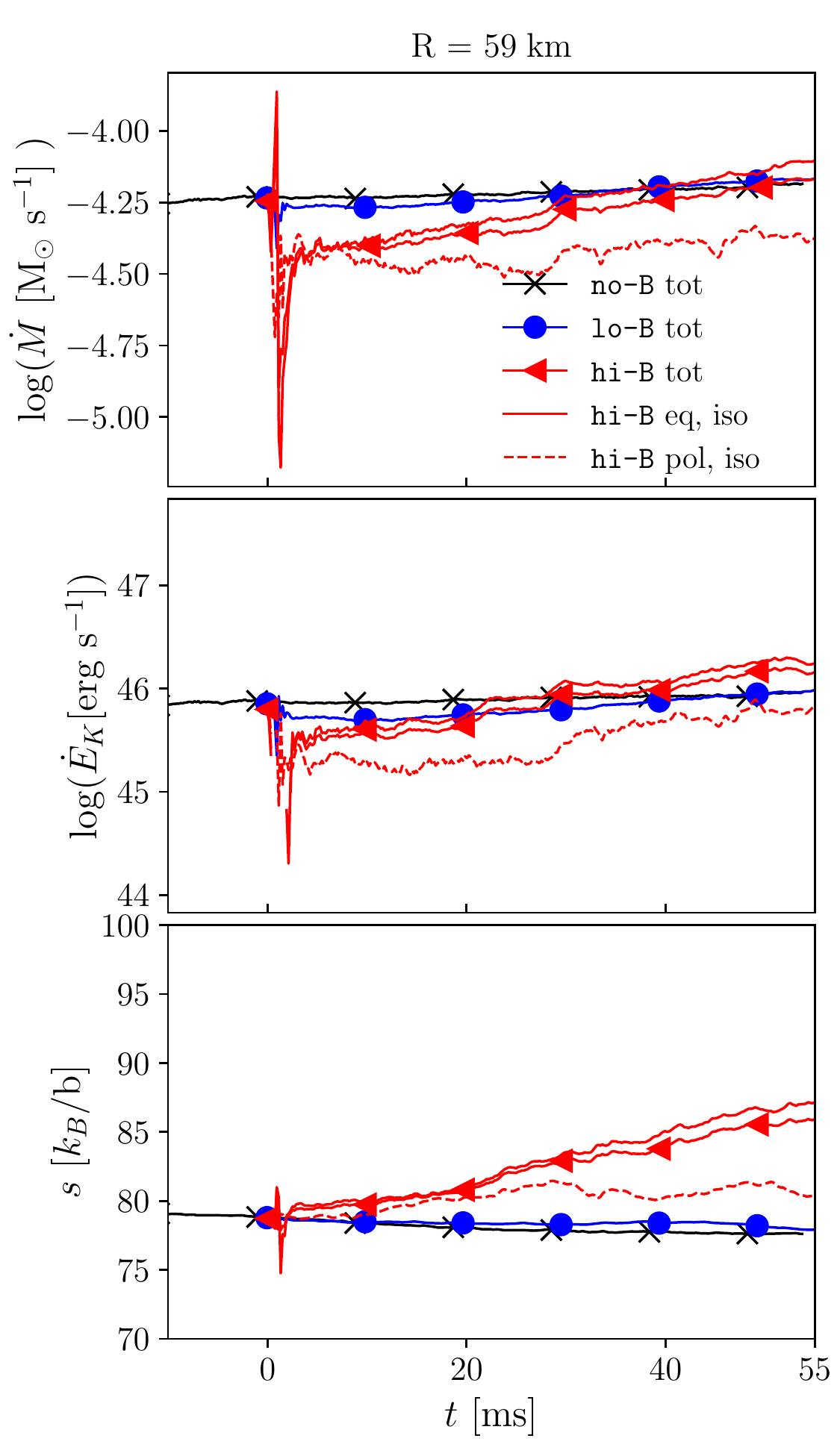}
\caption{Mass-loss rate $\dot M$ (top), kinetic power $E_K=(W -1) \dot{M}$ (center), where $W$ is the Lorentz factor, %$\frac12 \dot M v^2$ (where $v^2 =1-1/W^2 $ and $W$ is the relative lorentz factor between the fluid and eulerian frames)
and $\dot M$-weighted average entropy $s$ (bottom) of outflows through a spherical surface of radius $r=60$ km, as a function of time since magnetic field initialization.  For all models \texttt{no-B} (black, $\times$), \texttt{lo-B} (blue circles), and \texttt{hi-B} (red triangles) these quantities are shown across the full 4$\pi$ solid angle. For model \texttt{hi-B} (red), we show separately the isotropic equivalent mass-loss rate as well as kinetic power, and average-entropy across a solid angle at high latitudes close to the poles ($\theta = 0^\circ - 45^\circ$ and $\theta = 135^\circ - 180^\circ$; dotted line) and near the equatorial plane ($\theta = 45^\circ - 135^\circ$; solid line).}
\label{fig:mdot_s_evol}
\end{figure}

The mass-loss rate $\dot M$ through a fixed surface of radius $\approx 60$ km is also suppressed after the magnetic field is initialized ($t \approx 0$) by almost a factor of 2 from its original value, from $\dot M \approx 5.6\E{-5} M_{\odot} ~\mathrm{s}^{-1}$ to $\dot M \approx 3.2\E{-5} M_{\odot}~ \mathrm{s}^{-1}$ (top panel of Fig.~\ref{fig:mdot_s_evol}).  Dividing the outflow into separate polar ($\theta =0 - 45^\circ$, $135 - 180^\circ$) and equatorial ($\theta =45^\circ - 135^\circ$) angular sectors, we see that the isotropic-equivalent mass-loss rate $\dot{M}_{\rm iso}$ along the equatorial direction has returned to its original value by $t\approx 35$ ms and continues to increase, approaching $\dot{M}_{\rm iso} \approx 10^{-4} M_{\odot}~ \mathrm{s}^{-1}$ by $t \approx 55$ ms.  By contrast, $\dot{M}_{\rm iso}$ in the polar region remains significantly suppressed for the entirety of the simulation.  As we describe below, this suppression is the result of the extra work the polar outflows must perform to open field lines in this region and escape to infinity.

\begin{figure}
\centering
    \includegraphics[width=.4\textwidth]{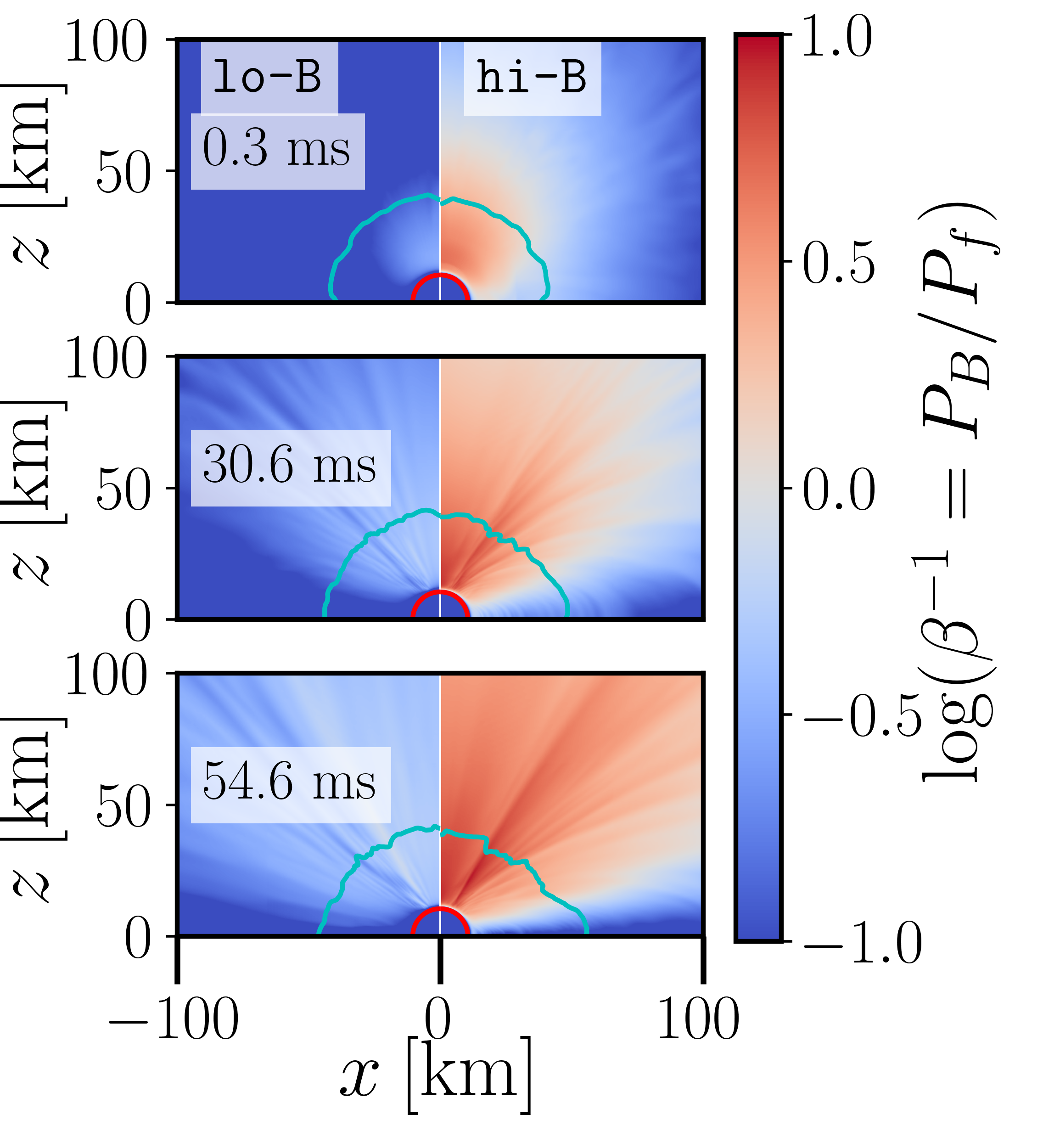}
\caption{Snapshots of the magnetic-to-fluid pressure ratio $\beta^{-1}$ in the $y=0$ magnetic dipole axis plane for models \texttt{lo-B} and \texttt{hi-B}. The top panel shows the moment the magnetic field is initialized ($t\approx 0$ ms), while the middle and bottom panels are at $t\approx 31$\,ms and $t\approx 55$\,ms, respectively.  A cyan contour represents the alpha particle formation surface ($X_\alpha=0.5$) around which $r$-process seed nuclei begin to form, while the red contour represents the neutrinosphere surface ($\tau_{\bar \nu_e}=1$).}
\label{fig:mhd_series}
\end{figure}

\begin{figure}
\centering
    \includegraphics[width=.4\textwidth]{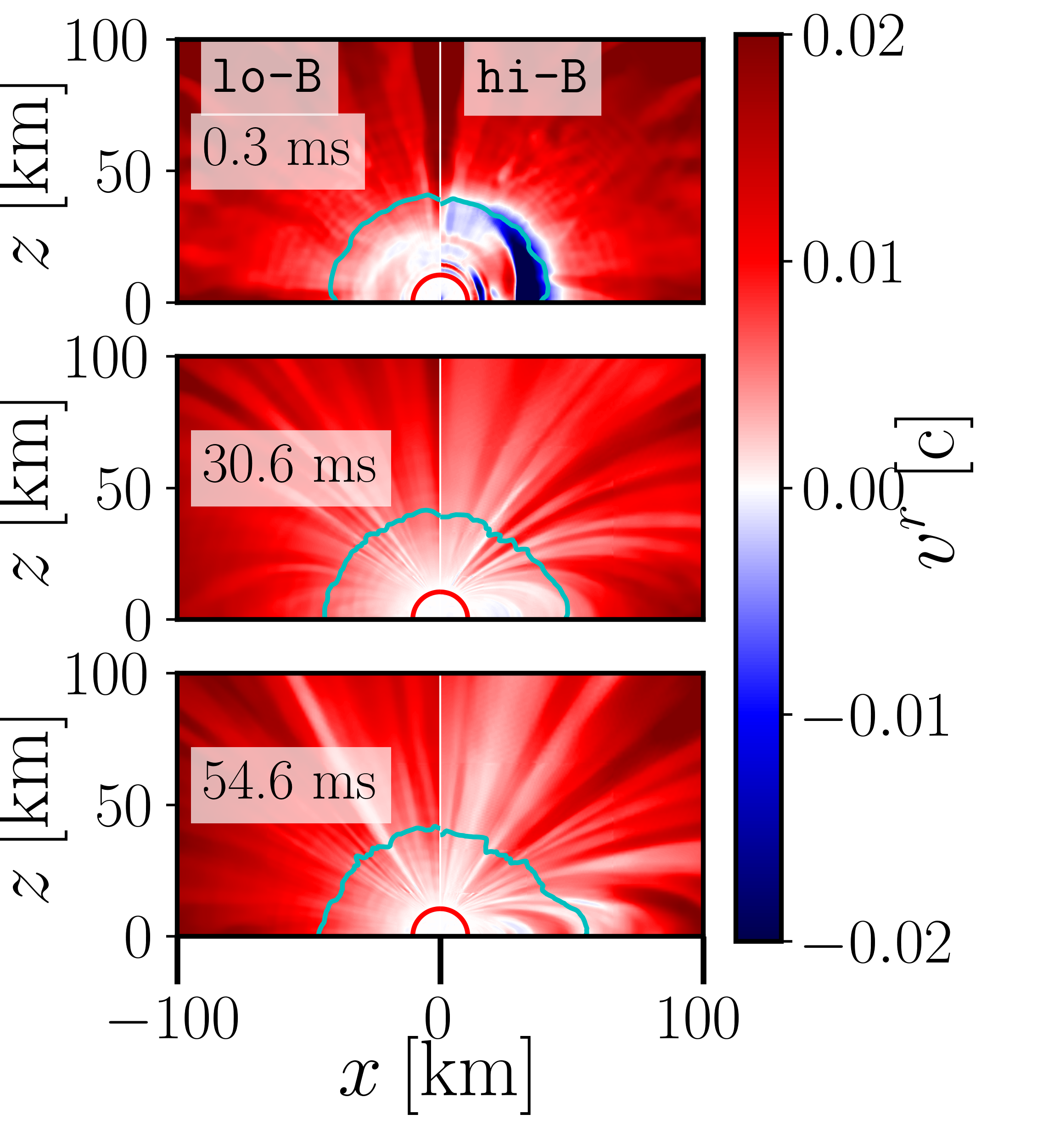}
\caption{Same as Fig.~\ref{fig:mhd_series}, but for radial velocity $v^r$.  The lower velocities of matter in the equatorial closed zone region, particularly in the first two snapshots, is apparent. }
\label{fig:vr_series}
\end{figure}

\begin{figure}
\centering
    \includegraphics[width=.4\textwidth]{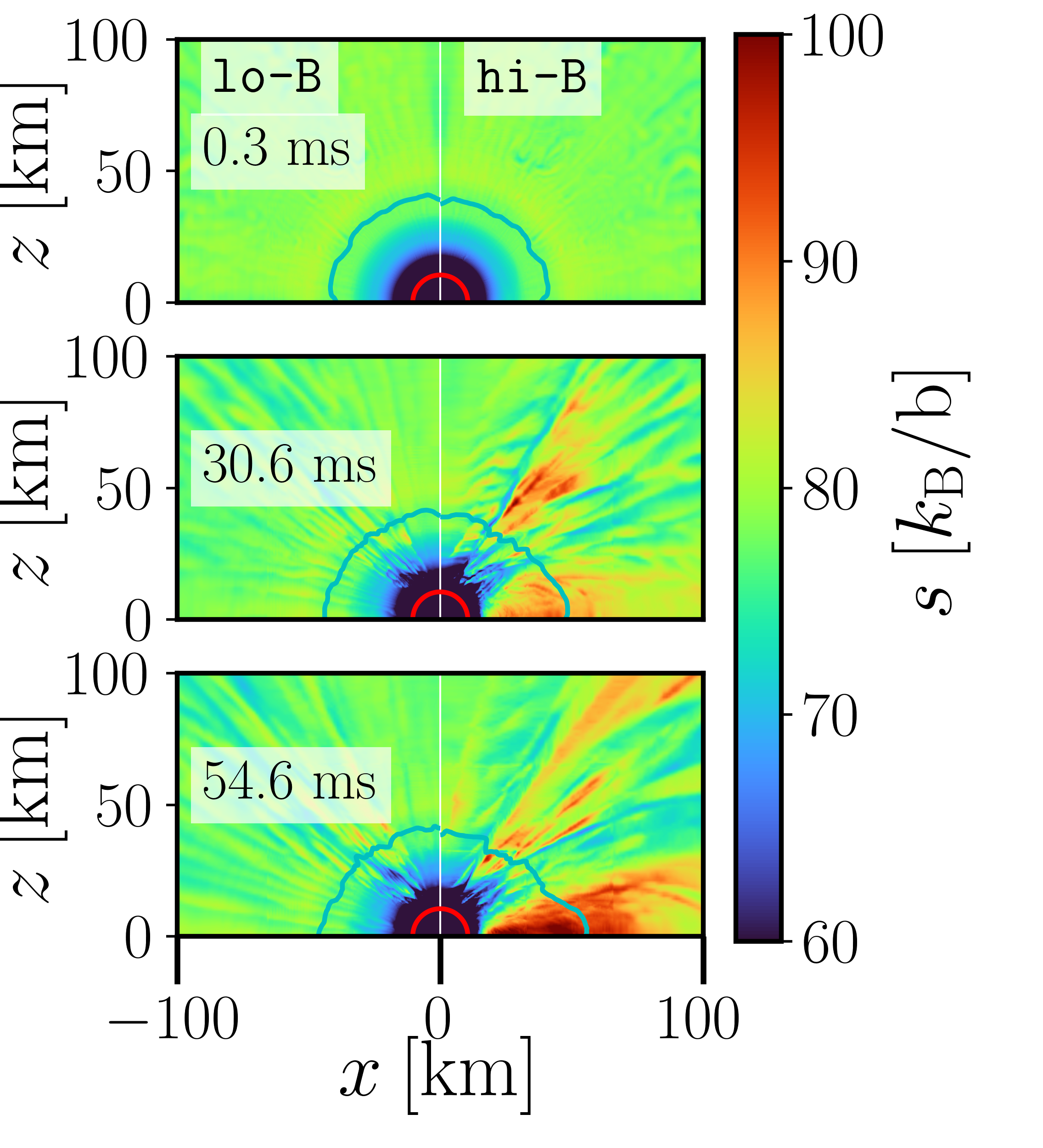}
\caption{Same as Fig.~\ref{fig:mhd_series}, but for specific entropy $s$.  The effect of the enhanced heating of matter in the equatorial closed zone is apparent.}
\label{fig:ent_series}
\end{figure}

As shown in Fig.~\ref{fig:vr_series}, the radial velocity remains small or negative $|v^r| \ll 0.01c$ in the equatorial region between $r \approx 15$ km and $r \approx 100$ km, indicating the presence of a sustained `trapped zone' at low latitudes.  The magnetic-to-fluid pressure ratio $\beta^{-1}$ begins to rise in the polar region and to drop in the equatorial region (Fig.~\ref{fig:mhd_series}). At the same time, the specific entropy in the $\theta=60^\circ-120^\circ$ equatorial trapped belt between $r=15$ km and $r=100$ km rises from $s\approx 75$ to $s \approx 100$ by $t \approx 55$ ms (Fig.~\ref{fig:ent_series}). 

\begin{figure}
\centering
    \includegraphics[width=.4\textwidth]{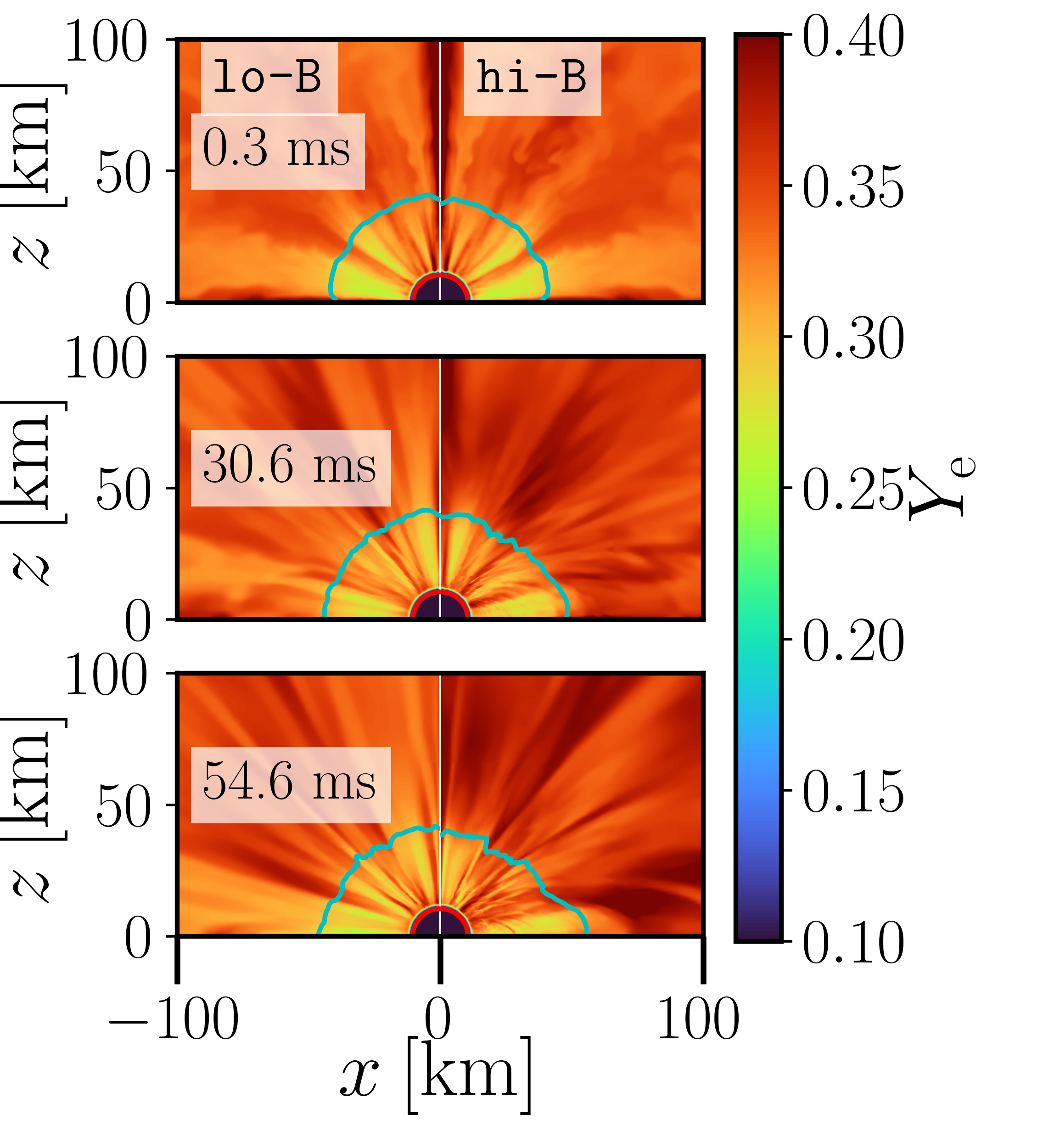}
\caption{Same as Fig.~\ref{fig:mhd_series}, but for $Y_e$.}
\label{fig:ye_series}
\end{figure}

\begin{figure}
\centering
    \includegraphics[width=.4\textwidth]{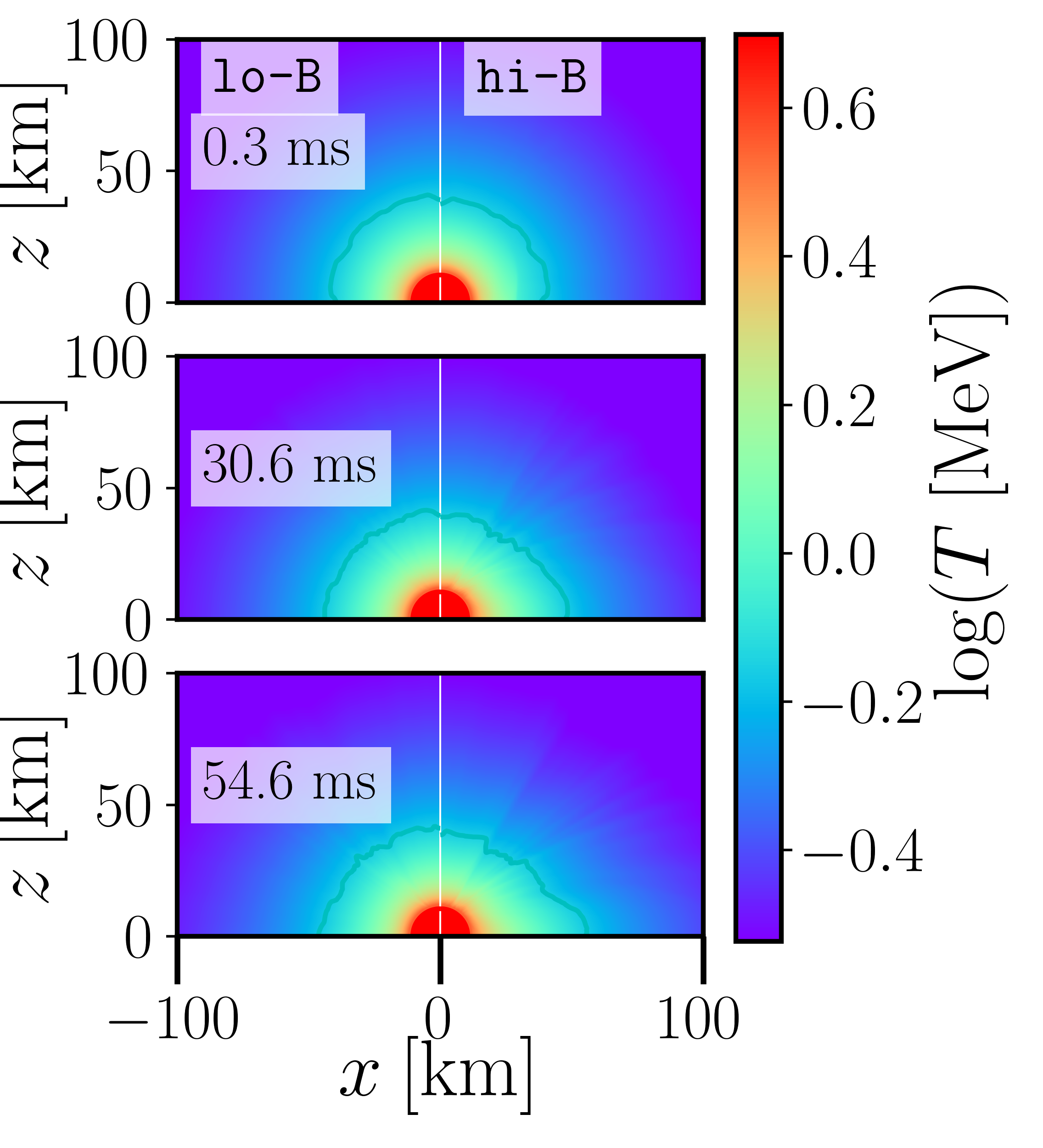}
\caption{Same as Fig.~\ref{fig:mhd_series}, but for temperature $T$.}
\label{fig:temp_series}
\end{figure}

Since magnetic fields do not strongly impact the hydrostatic structure of the PNS atmosphere near the neutrinosphere, neither the neutrino luminosities/energies nor the neutrinosphere radii are altered significantly by the magnetic field (Tab.~\ref{tab:models}). Insofar as the neutrino fluxes and energies determine the relative rate of $\nu_e$ and $\bar{\nu}_e$ absorption by the wind material, the wind's electron fraction is not altered considerably compared to the unmagnetized model (Fig.~\ref{fig:ye_series}).  The density and temperature profile, particularly of the inner hydrostatic atmosphere, also remain only mildly affected by the presence of the magnetic field (Fig.~\ref{fig:temp_series}).

% \begin{figure}
% \centering
%     \includegraphics[width=.4\textwidth]{vr_evol.png}
%     \includegraphics[width=.4\textwidth]{ent_evol.png}
%     \includegraphics[width=.4\textwidth]{mhd_evol.png}
%     \includegraphics[width=.4\textwidth]{Etot_evol.png}
% \caption{Cross-sectional profiles of wind quantities in a slice through the $y=0$ plane. From top to bottom: $v^r$, $s$, $p_B/p_f$, and $E_{\rm tot}$.\dkd{fix conflicting definitions of Etot.} The left and right panels of $v^r$ are at $t\approx 153$ ms and $t\approx 173$ ms, respectively. The left and right panels for all other quantities correspond to times $t\approx 178$ ms and $t\approx 208$ ms respectively. The magnetic field is initialized at $t=153$ ms. \dkd{put vr in separate fig? or not at all? otherwise confusing.}}
% \label{fig:progression}
% \end{figure}

\begin{figure}
\centering
    \includegraphics[width=.48\textwidth]{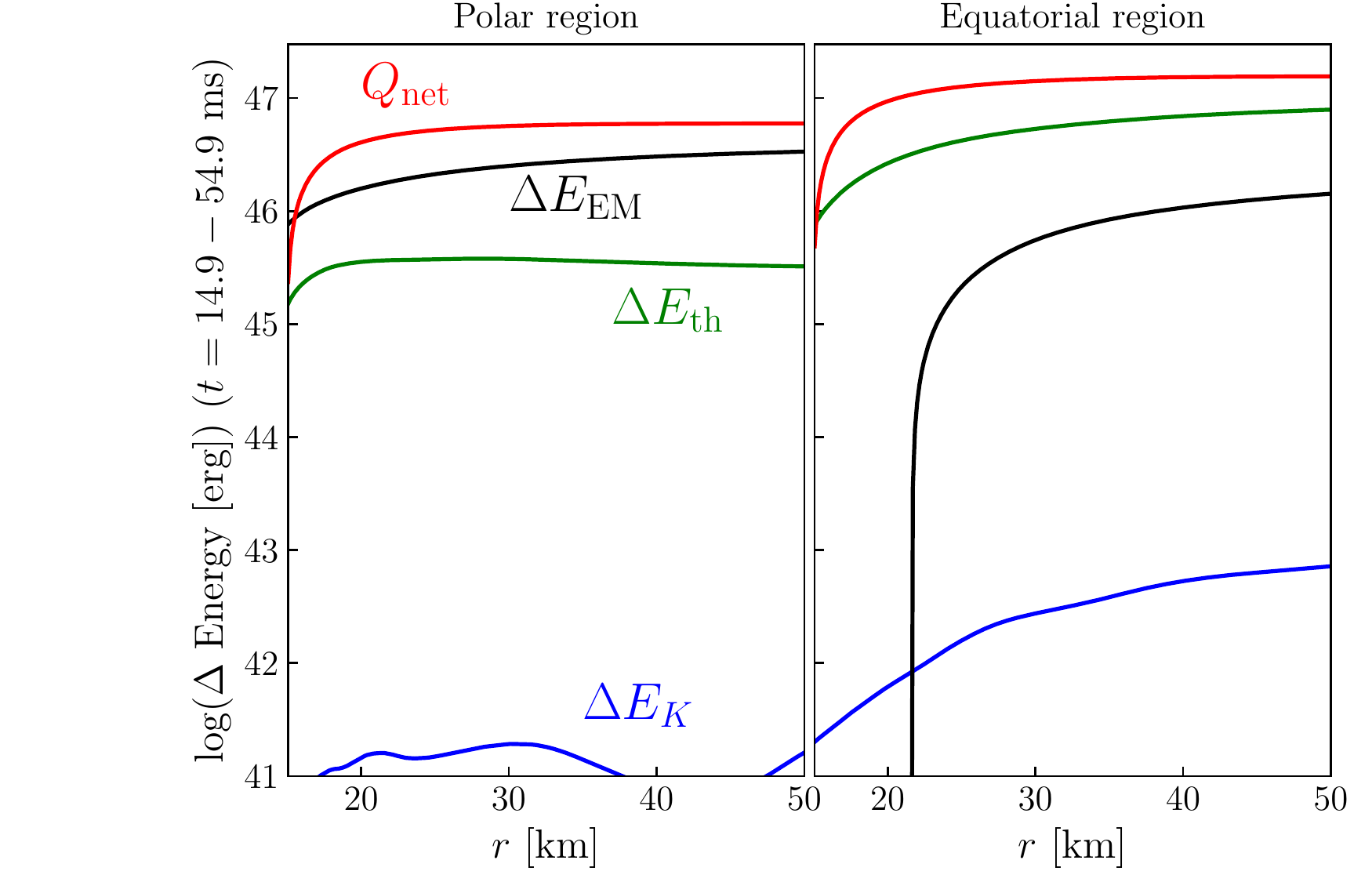}
\caption{Change in the energy contained within the volume extending from $r=15$ km to an outer radius $r$ between $t\approx 15$ ms and $t\approx 55$ ms following the initialization of the magnetic field for model \texttt{hi-B}. Left: Polar component, averaged azimuthally and over polar angles $\theta =0^\circ-45^\circ$ and $135^\circ-180^\circ$. Right: Equatorial component, averaged azimuthally and over polar angles $\theta =45^\circ-135^\circ$. The energies are defined as follows: kinetic energy $E_K=\int (W - 1)\rho W \sqrt{\gamma}\,d^3x$ (blue); thermal energy $E_{\rm th}=\int \epsilon \rho W \sqrt{\gamma}\,d^3x $ (green); electromagnetic energy $E_{\rm EM}=\int n_\mu n_\nu T_{\rm EM}^{\mu \nu} \sqrt{\gamma}\,d^3x$ (black); net neutrino heating per unit time $Q_{\rm net}=\int\int \dot q_{\rm net}\rho W \sqrt{\gamma}\,d^3x\, dt$ (red). Here, $T_{\rm EM}^{\mu \nu}$ is the electromagnetic stress-energy tensor, $n^\mu$ the 4-velocity of the Eulerian observer, $W$ the Lorentz factor associated with the total 4-velocity, and $\gamma$ the determinant of the 3-metric. The specific net neutrino heating rate $\dot q_{\rm net}$ is integrated over the time interval; all other quantities are evaluated at the final and initial times, with their difference plotted.}
\label{fig:energies}
\end{figure}

As discussed at the beginning of Sec.~\ref{sec:results}, the impact of the magnetic field on the wind dynamics can be understood in terms of the high magnetic-to-fluid pressure ratio in model \texttt{hi-B} (Fig.~\ref{fig:initial_mhd}, right panel). Outflowing matter in the polar regions tear open originally closed magnetic field lines, resulting in a split-monopole geometry at high latitudes (similar to that achieved across all outflow directions in model \texttt{lo-B}).  Given that the surface magnetic field strength is fixed, the transformation from $B \propto r^{-3}$ to $\propto r^{-2}$ by $\approx 30$ ms (Fig.~\ref{fig:Bpol}) causes the magnetic to fluid pressure ratio above the surface at high latitudes to increase with time.  

In addition, matter that would otherwise have traveled radially in the polar region is partially redirected along magnetic field lines, which bend towards lower latitudes.  This suppresses $\dot M_{\rm iso}$ in the polar region (Fig.~\ref{fig:mdot_s_evol}, top panel, dashed line). Field lines at mid-latitudes ($\theta \approx 30^\circ - 60^\circ$ and $120^\circ-150^\circ$) are gradually opened (`peeled off' from originally closed field lines at low latitudes) and the pressure ratio there similarly increases (Fig.~\ref{fig:mhd_series}). In the equatorial region ($\theta \approx 60^\circ-120^\circ$), however, magnetic tension remains high enough to oppose radial fluid motion, which is orthogonal to magnetic field lines in this region as a result of the dipole geometry.  The resulting `trapped zone' extends to $r \approx 50$\,km. 

To better understand the wind dynamics, Fig.~\ref{fig:energies} shows the cumulative change of various energies interior to a given radius above the PNS surface, over the $\approx$ 55 ms duration of the simulation, again broken down separately into polar (left panel) and equatorial angular sectors (right panel).  In both latitude ranges, the increase in the magnetic energy (black lines) exceeds the increases in the wind thermal (green lines) or kinetic (blue lines) energies.  This illustrates that most of the energy deposited by neutrino heating (red lines) is used to open magnetic field lines, rather than powering the wind.  The energy being expended to open field lines is not available to unbind matter from the gravitational potential well of the PNS and hence contributes to the initial suppression of the wind mass-loss rate and kinetic power shown in Fig.~\ref{fig:mdot_s_evol}.  Indeed, the magnetic energy rises with a greater delay in the equatorial belt relative to the polar regions, because it takes longer to open the closed magnetic field lines at lower latitudes (which are initially oriented perpendicular to fluid flow) through neutrino heating.  This is counterbalanced by a greater fraction of the neutrino heating rate being deposited into thermal energy in the equatorial region (the ratio of the black to red lines in Fig.~\ref{fig:mdot_s_evol}) compared to the polar region, because of the longer residence time of the matter trapped in the closed portion of the magnetosphere compared to the expansion time of a continuous outflow.

\begin{figure}
\centering
    \includegraphics[width=.4\textwidth]{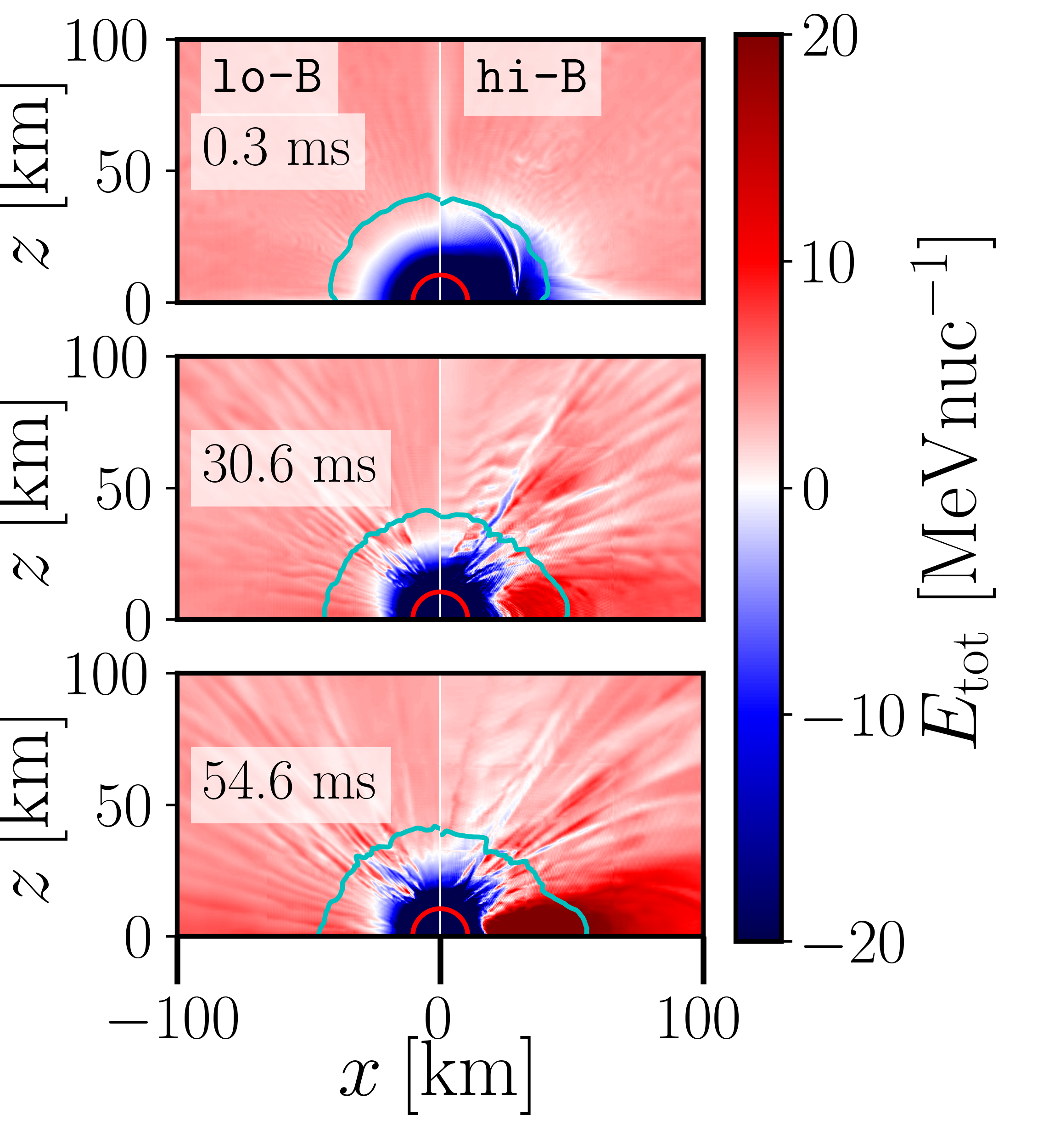}
    %\vspace{mm}
\caption{Same as Fig.~\ref{fig:mhd_series}, but for total specific energy $E_{\rm tot}$, including the effects of magnetic tension (Eq.~\ref{eq:Etot}).  The energy grows in the equatorial belt due to neutrino heating of matter trapped in the closed zone, becoming positive by the final snapshots across a greater region extending to smaller radii approaching the PNS surface.  As this high-entropy matter expands through the seed formation region (outside the cyan contour) it can potentially generate the conditions necessary for 2nd or 3rd-peak $r$-process nucleosynthesis.  }
\label{fig:etot}
\end{figure}

\begin{figure}
\centering
%\begin{noverticalspace}
\includegraphics[width=.49\textwidth]{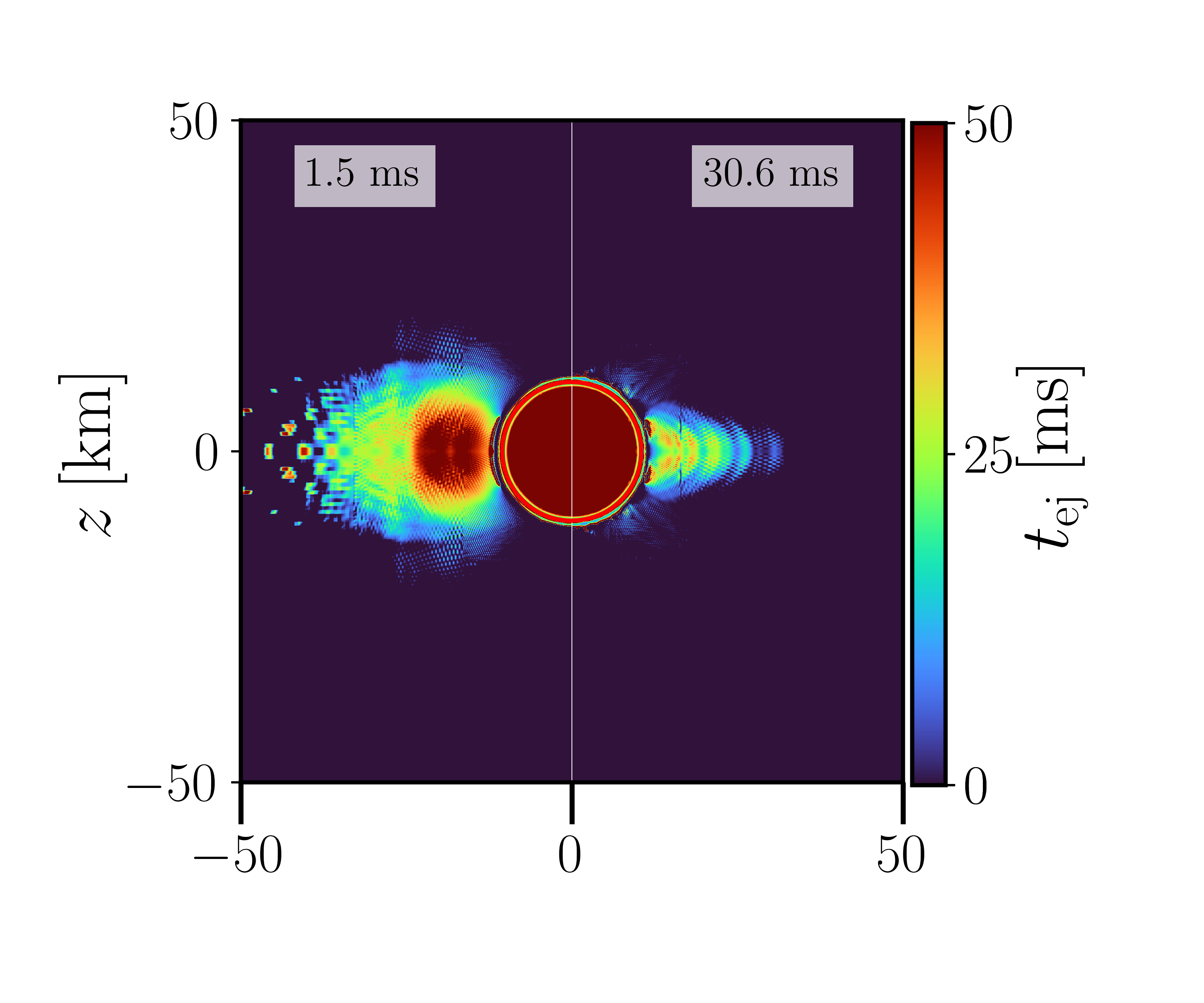}
\includegraphics[width=.49\textwidth]{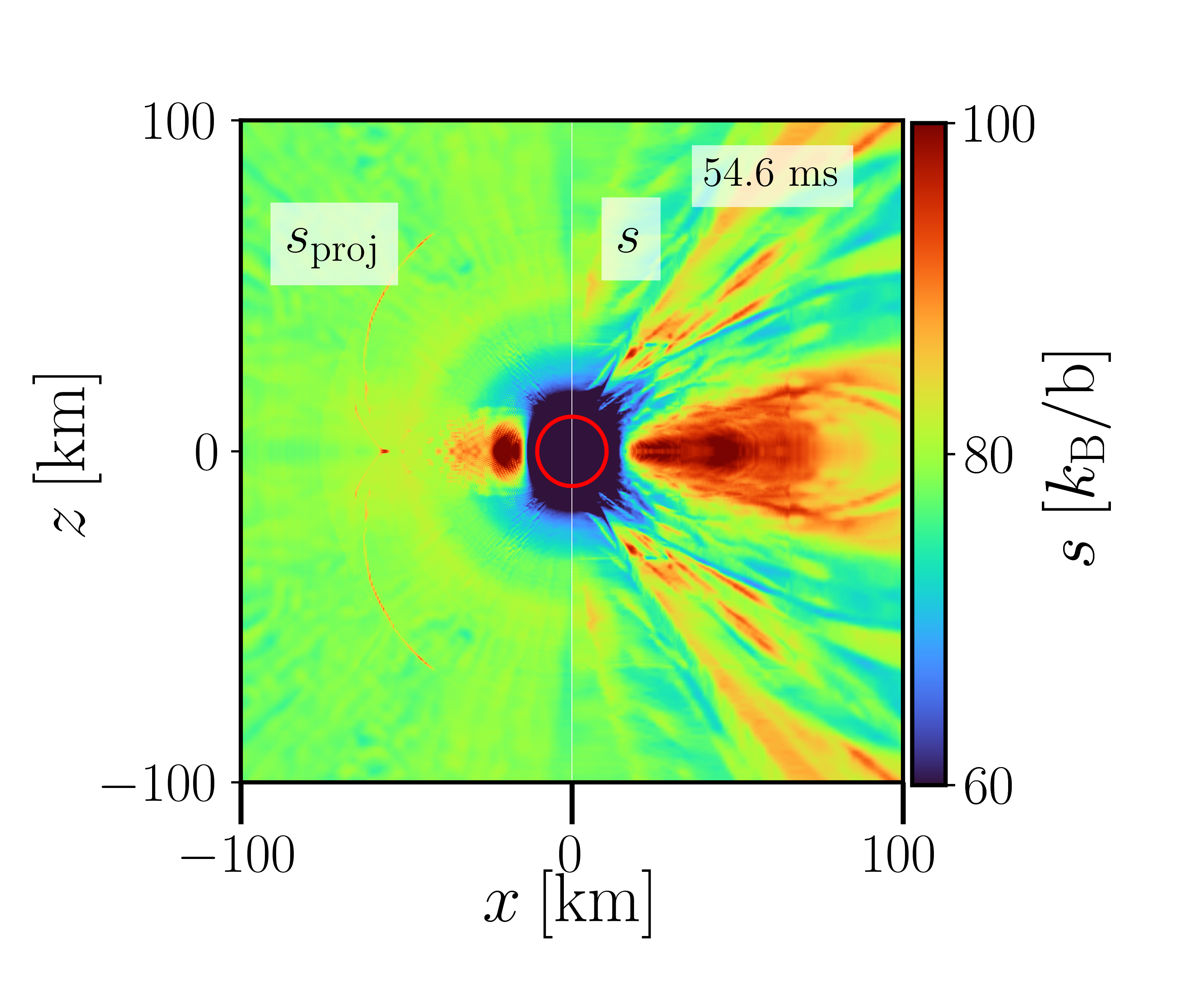}
%\end{noverticalspace}
\caption{Top: Closed-zone ejection timescale $t_{\rm ej}$ (Eq.~\ref{eq:tej}) computed just after magnetic field initialization (left) and 30\,ms later (right) for model \texttt{hi-B} in a slice through the $y=0$ plane. Bottom: Snapshots from model \texttt{hi-B} showing (left) the final entropy $s_{\rm proj}$ of the closed zone material achieved after a time $t_{\rm ej}$ (Eq.~\ref{eq:tej} from \citealt{Thompson03} applied to the initial snapshot) in comparison to (right) the actual entropy $s$ from the simulation at $t \approx 55$ ms after magnetic field initialization. The red contour represents the neutrinosphere surface ($\tau_{\bar \nu_e}=1$). The analytic estimate is roughly consistent with the actual entropy achieved in the trapped zone of the wind.} 
\label{fig:t_ej}
\end{figure}

The trapped zone is further illustrated by Fig.~\ref{fig:etot}, which shows the total specific energy of the fluid, 
\be
    E_{\rm tot}=-h' u_t-1, \label{eq:Etot}
\ee
where we have now modified the specific enthalpy $h$ to include the effects of magnetic tension according to
\be
 h' = 1+\epsilon +\frac{P_f-P_{B,\rm T}}{\rho}. \label{eq:h}
\ee
Here, $u_t$ is the $0$-component of the four-velocity, $\epsilon$ is the specific internal energy, and
\be
P_{B,\rm T} \equiv \sin \alpha \frac{B^2}{4\pi}\left( \frac{R_\nu}{R_{c}(r,\theta)}\right),
\label{eq:uB}
\ee
is the effective confining pressure of magnetic tension, where $\alpha$ is the local angle between the radial direction and the magnetic field line direction and $R_{c}$ is the radius of curvature. The bound region $E_{\rm tot} < 0$ under this definition extends to 50\,km in equatorial regions for model \texttt{hi-B}, consistent with the geometry of a trapped zone evident in the velocity field as discussed above (Fig.~\ref{fig:vr_series}).

Since the neutrino properties are not significantly altered by the strong magnetic field, the gain region remains almost identical to that in models \texttt{no-B} and \texttt{lo-B}, starting at the PNS surface and extending out to ${\approx}50\,{\rm km}$.  As a result, the trapped zone experiences additional neutrino heating, increasing the ratio of fluid pressure to magnetic tension pressure  (Fig.~\ref{fig:mhd_series}).  Unlike for the quasi steady-wind solutions achieved in models \texttt{no-B} and \texttt{lo-B}, the specific entropy in the trapped magnetosphere, $\Delta s=\int \dot{q}_{\rm net}/T dt$, thus rises monotonically with time (Fig.~\ref{fig:ent_series}).

Eventually, once the trapped zone is heated sufficiently for the fluid pressure to exceed the magnetic pressure, the field lines should open and the high-entropy matter will be ejected.  The timescale over which this occurs, $t_{\rm ej}$, can be estimated by (e.g.~\citealt{Thompson03,thompson_high-entropy_2018})
\be
t_{\rm ej} = \frac{P_{B,\rm T}-P_f}{\dot{q}_{\rm net} \rho}.
\label{eq:tej}
\ee

The top panel of Fig.~\ref{fig:t_ej} shows the ejection time $t_{\rm ej}$ (Eq.~\ref{eq:tej}) computed for model \texttt{hi-B} at two times: just after the magnetic field is initialized (left panel) and at a snapshot taken $\approx$31 ms later (right panel).  Initially, $t_{\rm ej}$ peaks in the equatorial belt above the PNS surface at a value $\simeq 50$\,ms.  By $\approx 31$\,ms the trapped region has shrunk in size, and the maximum ejection time has dropped to $t_{\rm ej}\simeq 20$ ms, roughly as expected given the amount of time elapsed.  The entropy of the trapped zone material at the time of ejection $t=t_{\rm ej}$ can be estimated as (e.g. \citealt{Thompson03})
\be
s_{\rm proj} \approx s +t_{\rm ej}\dot q_{\rm net}/T,
\label{eq:s_proj}
\ee
where $t_{\rm ej}\dot q_{\rm net}/T$ is the projected entropy gain at the approximate time of ejection (Eq.~\ref{eq:tej}).  The bottom panel of Fig.~\ref{fig:t_ej} compares this future-projected final entropy $s_{\rm proj}$ at the time of magnetic field initialization (left panel) to the actual entropy achieved $t_{\rm ej} \approx 55$ ms later (right panel), around when trapped zone material is expected to be ejected. The rough quantitative agreement between the projected and achieved entropy suggests that ejection of the closed zone is imminent near the end of our simulation. 

Unlike the equatorial region, the polar region is not trapped by the magnetic field; thus the material there does not experience additional heating and the entropy of the outflowing material is similar to that obtained in the weakly magnetized models \texttt{lo-B} and \texttt{no-B} (Fig.~\ref{fig:ent_series}).  The bottom panel of Fig.~\ref{fig:mdot_s_evol} shows for model \texttt{hi-B} how the entropy in the equatorial region at radius $\approx 60$ km grows in time (solid red line, bottom panel) in comparison to the roughly constant entropy of \texttt{lo-B} and \texttt{no-B} (blue and black lines) and for polar outflows in model \texttt{hi-B}.  Given the neutrino luminosities/energies and the strength of the magnetic field of our simulations, the entropy gain we find agrees with that found by \citet{thompson_high-entropy_2018} (see their Fig.~5).

\begin{figure*}
\centering
\includegraphics[width=0.5\textwidth]{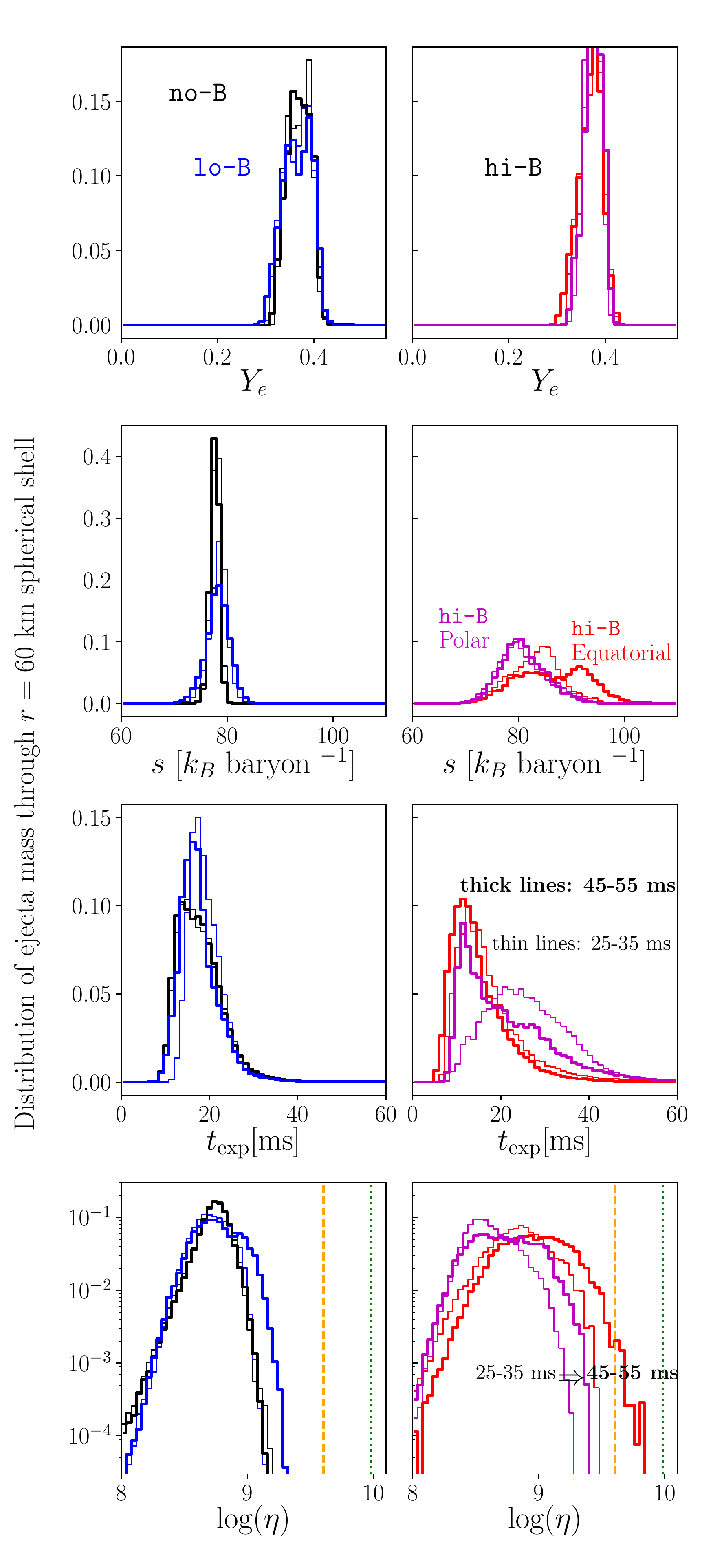}
\caption{Histograms of various quantities relevant to $r$-process nucleosynthesis as measured through a 60 km spherical shell.  The left panel shows the unmagnetized (\texttt{no-B}, black) and weakly magnetized (\texttt{lo-B}, blue) models, while the strongly magnetized model (\texttt{hi-B}) is shown on the right, broken down separately into the polar (purple) and equatorial (red) outflows, respectively (defined over the same angular domains as in Fig.~\ref{fig:mdot_s_evol}).  For the unmagnetized models we show results over the full time interval 25--55\,ms, because the outflow is approximately time-stationary, while for the magnetized models we separately bin results for 25--35\,ms (thin lines) and 45--55\,ms (thick lines). At a given time, quantities are weighted pointwise on the spherical grid by the local mass outflow, they are binned and normalized by the total mass outflow through the entire surface over the time interval. We approximate the outflow expansion time (Eq.~\ref{eq:texp}) as $t_{\rm exp} = R/v(R)$, where $v(R)$ is the total wind speed measured at the shell. Threshold values for $\eta$ (Eq.~\ref{eq:eta}) required for neutron captures to reach the 2nd (yellow dashed vertical line) and 3rd (green dotted vertical line) $r$-process peaks \citep{Hoffman+97} are indicated for comparison.}
\label{fig:hist}
\end{figure*}

Figure~\ref{fig:ent_series} shows that the alpha-particle formation surface (cyan contour) passes through the equatorial trapped zone region.  \citet{Thompson&udDoula18} found that neutrino heating of material in the trapped zone increases the ratio of gas to magnetic pressure until a minimum $\beta^{-1}$ is reached, after which the closed zone is ejected as a single coherent structure.  In our \texttt{hi-B} model we find that material from the trapped zone already has begun to leak out by $t \gtrsim 35$ ms.  We speculate that the energy released by alpha-particle recombination, neglected by \citealt{Thompson&udDoula18}, contributes to unbinding matter from the edge of the closed zone in our simulation, in addition to neutrino heating.

\subsection{Implications for $r$-process nucleosynthesis}
\label{sec:nucleosynthesis}

Figure~\ref{fig:hist} shows angle- and time-sampled histograms of the outflow properties relevant to $r$-process nucleosynthesis as measured through a spherical surface of radius $\approx$ 60 km for models \texttt{no-B} and \texttt{lo-B} (left column).  We also show results for model \texttt{hi-B} (right column), in this case broken down separately into polar and equatorial outflows and (since the highly magnetized case has not reached a steady state) shown in separate time intervals $25-35$ ms and $45-55$ ms after activation of the magnetic field, respectively.  Some of these time-averaged wind properties are also summarized in Tab.~\ref{tab:wind_props}.

The $Y_e$ distributions for all models are isotropic and nearly identical to one another (see also Fig.~\ref{fig:ye_series}); this is as expected because the outflow speeds are sufficiently low that neutrino absorptions have time to bring $Y_e$ into equilibrium, and as already mentioned, the properties of the neutrino radiation are similar between the magnetized and unmagnetized models (Tab.~\ref{tab:models}).  The entropy distributions for the weakly magnetized models \texttt{lo-B} and \texttt{no-B} are nearly identical to each other, centered around $s \approx 78$ with a relatively narrow spread of $\Delta s \approx \pm 1$ (Tab.~\ref{tab:wind_props}).  This is also as expected given the previously noted similarities between the unmagnetized and weakly magnetized models, for both of which the outflow reaches an approximate steady-state (Sec.~\ref{sec:weakmag}).

In the strongly magnetized model \texttt{hi-B}, the entropy distribution of the polar outflows overlaps that of the weakly magnetized case, though with a significantly larger spread $\Delta s \approx \pm 3.6$.  By contrast, the mean entropy of the equatorial \texttt{hi-B} outflows is shifted to a higher value $s \gtrsim 80$, with an even greater spread $\Delta s \approx \pm 6$ in the distribution.  Furthermore, the mean entropy rises significantly in time (compare the 25--35\,ms vs.~45--55\,ms samples in Fig.~\ref{fig:hist}), such that by the end of the simulation values as high as $s \gtrsim 100$ are achieved (see also Fig.~\ref{fig:ent_series}). 

The expansion timescale distribution for models \texttt{lo-B} and \texttt{no-B} are again nearly identical, centered around $\sim 20$ ms. For both polar and equatorial ejecta in model \texttt{hi-B} the distribution extends to larger expansion times than the weakly magnetized cases, due to the significant trapping effect of the magnetic field.  The nearly indistinguishable distributions in $Y_e$, $s$, and $t_{\rm exp}$ between models \texttt{lo-B} and \texttt{no-B} imply that the distribution of $\eta$ (Eq.~\ref{eq:eta}) should also agree; this is seen in the bottom panel of Fig.~\ref{fig:hist}. As already noted (Sec.~\ref{sec:weakmag}), both models remain well below the required threshold $\eta$ for 2nd-peak $r$-process production. 

By contrast, for the strongly magnetized model \texttt{hi-B}, the $\eta$ distribution of the equatorial outflows extend to higher values due to the higher entropy ($\eta \propto s^{3}$).  Over the course of the simulation, $\eta$ increases from a mean value of $\approx 4\E{8}$ to $\approx 1\E{9}$, with the high-$\eta$ tail (about $0.4\%$ of equatorial material) achieving values $\gtrsim 4\E{9}$ necessary for 2nd peak $r$-process production $\approx 45-55$ ms after the magnetic field is initialized.  We conclude that$-$all else being equal (e.g., in terms of their neutrino emission properties)$-$strongly magnetized PNS are more promising $r$-process sources than weakly magnetized PNS.

The outflow entropies for our model \texttt{hi-B} are broadly consistent with those found by \citet{Thompson&udDoula18} using 2D axisymmetric MHD simulations, for roughly the same surface magnetic field strength.  These authors also found that a small fraction of the ejecta reaches large values of $\eta \gtrsim 10^{10}$, sufficient for a 2nd or even 3rd peak $r$-process, due to the transient ejection of high-entropy matter from the closed zone.  Although the evolution of our \texttt{hi-B} model indeed resembles a single episode of closed-zone inflation and eruption, our simulations unfortunately cannot be run as long as those by \citet{Thompson&udDoula18,Prasanna+22} due to the higher computational cost of our 3D GRMHD simulations that aim to marginally resolve the neutrinosphere, versus axisymmetric 2D simulations.  

Beyond their computational cost, the duration of our simulations are also limited by numerical issues: at late times $t\gtrsim 60$ ms, spurious violations of $\nabla\cdot\mathbf{B}=0$ at refinement level boundaries of our fixed Cartesian grid hierarchy of concentric boxes at a level of $\sim\!1\%$ have accumulated due to interpolation operations over a total of $\gtrsim\!210$\,ms of evolution and residual violations introduced by initializing a large-scale dipole magnetic field. At this level, we do not entirely trust subsequent results and choose not to consider those data in our analyses, even though the conditions for heavy $r$-process nucleosynthesis are seen to be improving with time as higher entropy material expands through the seed formation region.  We refer to Appendix \ref{appendix:divB} for a more detailed discussion of the issue of $\nabla\cdot\mathbf{B}=0$ violation.

%divergence constraint violations appear in our GRMHD simulations, which in general do not occur in non-relativistic ideal-MHD codes such as the one used in \citet{Thompson&udDoula18}.  A future version of the GRMHD code which employs the vector potential formalism (where $\nabla \cdot \vec{B}$ conservation on refinement boundaries is intrinsic to the evolution) will allow us to evolve the system to longer time scales.

%As a result we are not able to simulate the wind over longer timescales.\dkd{seems like not the best excuse...$\beta^{-1}$ is high in the polar region, but divB primarily violated in equatorial region.} A future version of the GRMHD code which uses the vector potential formalism (where $\nabla \cdot \vec{B}$ conservation on refinement boundaries is intrinsic to the evolution) will allow us to evolve the system to longer time scales.

\section{Summary and Conclusions}
\label{sec:conclusions}

 We have performed 3D GRMHD simulations including M0 neutrino transport of magnetized PNS winds to explore the impact that magnetar-strength dipole surface magnetic fields have on the outflow properties, with a particular focus on the conditions necessary for a successful $r$-process via the $\alpha$-rich freeze-out mechanism in mildly neutron-rich winds.  Our results can be summarized as follows.

\begin{itemize}
    \item For even the strongest magnetic fields that we consider ($B_{\rm S} = 2.5\times 10^{15}$ G; model \texttt{hi-B}), magnetic forces do not appreciably impact the hydrostatic structure of the wind near the neutrinosphere radii.  As a result, the properties of the neutrino radiation ($L_\nu$, $E_\nu$, $R_\nu$) which dictate the equilibrium electron fraction and specific heating rate in the gain region, are similar between the magnetized and unmagnetized models.
    
    \item In the case of a relatively weak magnetic field ($B_{\rm S} \simeq 6.1\times 10^{14}$ G; $B_{\rm S} < B_{\rm crit}$; model \texttt{lo-B}) for which $\beta^{-1} = P_{\rm B}/P_{\rm f} \lesssim 1$, the dipole field structure is torn open by neutrino-driven outflows within $\sim$ 10 ms, and the magnetic field takes on a split-monopole configuration by $\approx$40 ms (Figs.~\ref{fig:split_mono}, \ref{fig:Bpol}).  Outflow properties such as the mass-loss rate and entropy are approximately spherical and quantitatively similar to those from the otherwise similar unmagnetized PNS model \texttt{no-B} (e.g., Figs.~\ref{fig:no-B}, \ref{fig:hist}).
    
    \item In stark contrast, the wind structure of the highly magnetized model ($B_{\rm S} \simeq 2.5\times 10^{15}$ G $B_{\rm S} > B_{\rm crit}$; model \texttt{hi-B}) differs qualitatively from the weakly magnetized cases.  Outflows that emerge along the polar axis of the dipole follow open magnetic field lines and are broadly similar in their isotropic-equivalent properties to the spherical unmagnetized and weakly magnetized winds.  One exception is the isotropic mass-loss rate, which is initially suppressed compared to a weakly magnetized wind, because a significant portion of the energy deposition from neutrino heating goes into opening polar magnetic field lines rather than lifting matter out of the gravitational potential of the star (Fig.~\ref{fig:energies}).  
    
    By contrast, outflowing material in the equatorial regions of the wind are initially trapped by the non-radial magnetic field at lower latitudes (Figs.~\ref{fig:mdot_s_evol}, \ref{fig:vr_series}, \ref{fig:t_ej}), with the magnetosphere in this region maintaining a dipole field structure well above the PNS surface (Fig.~\ref{fig:split_mono}).  Neutrino heating raises the thermal pressure of the trapped fluid in the equatorial region until it obeys $P_f > P_B$ \citep{Thompson03,Prasanna+22}, at which point fluid begins to escape and the closed zone begins to shrink from the outside inwards.  Energy input from $\alpha$-particle formation appears to aid the ejection of matter from the equatorial regions, and by the end of the simulation the isotropic-equivalent mass-loss rate even slightly overshoots that of the otherwise equivalent unmagnetized wind (Fig.~\ref{fig:mdot_s_evol}).  
    
    \item The weakly magnetized wind model achieves a rough steady-state and does not show significant entropy growth relative to the unmagnetized model, because matter is not trapped by the magnetic field (Fig.~\ref{fig:no-B}, top panel; Fig.~\ref{fig:mdot_s_evol}, bottom panel). 
    By contrast, plasma trapped in the strongly magnetized model causes the mean entropy of the trapped and eventually outflowing material from the equatorial region to rise, its standard deviation grows concurrently (Fig.~\ref{fig:hist}, second row, right panel), over the course of $\sim 50-60$ ms.  The mean expansion time of the equatorial outflows through the seed formation region is also moderately larger compared to the weakly magnetized cases because of the suppressed outflow speed. 
    
    \item For the strongly magnetized model, the heating profile and magnetic field strength in the trapped equatorial belt imply an ejection timescale of the trapped plasma of $\sim 50$ ms, following the analytic estimates of \citet{thompson_high-entropy_2018}  (Fig.~\ref{fig:t_ej}); although we do not see a discrete ejection event, a continuous slow but accelerating ``peeling'' of the trapped zone is observed to occur on this timescale. The projected entropy gain \citep{Thompson03} broadly agrees with the rise in entropy we observe in the simulation. 
    %\dms{Perhaps comment on continuous leakage due to release of energy from alpha-particle formation, depending on whether we find this to be a significant effect.}
    
    \item The $r$-process figure-of-merit parameter $\eta$ for unmagnetized and weakly magnetized models are similar ($\lesssim 10^9$), remaining well below the required threshold ($\approx 4 \E{9}$) to produce 2nd peak $r$-process elements (Fig.~\ref{fig:hist}). By contrast in the strongly magnetized model, due to the monotonic rise in the mean entropy of the equatorial outflows (particularly a ``tail'' of matter extending to high entropy $s \gtrsim 100$), sufficiently high $\eta$ may be achieved for a small subset ($\approx 0.4\%$) of equatorial material, within $\sim $50 ms of magnetic field initialization.  Following this trend to later times than the duration of our simulation, we conclude that a moderate fraction of the time-averaged wind material could well attain values of $\eta$ that surpass the 2nd and potentially also 3rd $r$-process peaks. Though due to numerical limitations we cannot follow the multiple cycles of trapped-zone inflation and mass ejection seen by \citet{Thompson&udDoula18,Prasanna+22}, our results are in broad agreement with the findings of these authors.
\end{itemize}

Paper I demonstrated that rapid rotation in unmagnetized PNS winds tends to reduce the entropy of neutrino-driven outflows, while in the present paper we have shown that a strong magnetic field tends to increase the wind entropy.  Although some aspects of the phenomena we have studied will be ``additive'' (i.e., neutrino-heating driven ejections from a rotating magnetosphere), qualitatively new features of the wind properties, such as magneto-centrifugal acceleration, are expected to emerge through the combined impact of rapid rotation and strong magnetic fields (e.g., \citealt{Thompson+04,Metzger+07,Vlasov+17,Prasanna+22,Combi&Siegel23,Raives+23,Prasanna+23}).  Rotating proto-magnetar winds will be the focus of future work.

\vspace{0.2cm}

\begin{acknowledgements}
We thank Erik Schnetter, Roland Haas, and Luciano Combi for discussions and support. This research was enabled in part by support provided by SciNet (www.scinethpc.ca) and Compute Canada (www.computecanada.ca). The authors gratefully acknowledge the computing time granted by the Resource Allocation Board and provided on the supercomputer Lise and Emmy at NHR@ZIB and NHR@Göttingen as part of the NHR infrastructure. The calculations for this research were conducted with computing resources under the project mvp00022. DKD and BDM acknowledge support from the National Science Foundation (grant \#AST-2002577). DMS acknowledges the support of the Natural Sciences and Engineering Research Council of Canada (NSERC), funding reference number RGPIN-2019-04684. Research at Perimeter Institute is supported in part by the Government of Canada through the Department of Innovation, Science and Economic Development Canada and by the Province of Ontario through the Ministry of Colleges and Universities.

\software{The Einstein Toolkit (\citealt{Loffler+12}; \href{http://einsteintoolkit.org}{http://einsteintoolkit.org}), GRMHD\_con2prim \citep{siegel_grmhd_con2prim_2018}, PyCactus \citep{Kastaun21}, Matplotlib \citep{matplotlib07}, NumPy \citep{numpy20}, SciPy \citep{scipy20}, and hdf5 \citep{hdf5}.}
\end{acknowledgements}
\bibliographystyle{aasjournal}
\bibliography{refsPNS,Relastro_PI_UoG}

\appendix
\section{Numerical Tests}
\label{appendix:numerical_details}

This Appendix presents a number of tests we have performed on the simulation results, which justify simplifications made in the grid setup and assumptions regarding the microphysics we have included.

\subsection{Spatial Resolution}
\label{appendix:resolution}

\begin{figure}
\centering
    \includegraphics[width=0.49\textwidth]{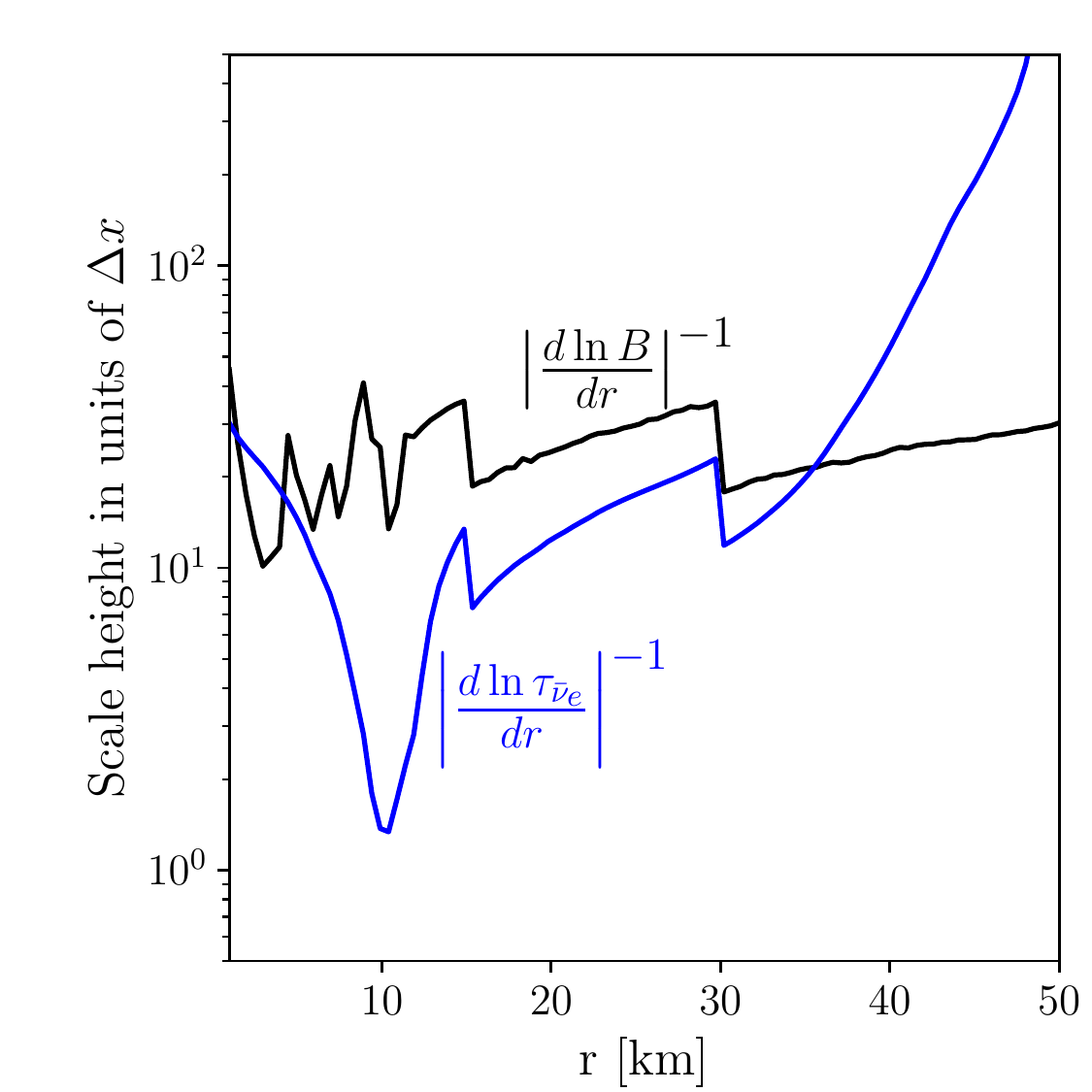}
    \caption{Vertical scale-height as a function of radius of the magnetic field $B$ (black) and neutrino optical depth $\tau_{\bar \nu_e}$ (blue) for model \texttt{hi-B}, normalized by the grid step size $\Delta x$.  The radial profiles correspond to the slice through the magnetic dipole axis ($y=0$ plane), averaged over polar angle between $\theta =0^\circ$ and $\theta=45^\circ$, and time-averaged over the first 4 ms after the magnetic field is initialized. Discontinuities at 15\,km and 30\,km reflect refinement level boundaries.}
\label{fig:mag_scale_height}
\end{figure}

Figure~\ref{fig:mag_scale_height} shows the vertical scale-height of the magnetic field strength $B$ and neutrino optical depth $\tau_{\bar \nu_e}$ as a function of radius for the strongly magnetized model \texttt{hi-B}. We resolve the magnetic field by at least 10 grid points throughout the entire simulation domain. The same conclusion holds for model \texttt{lo-B}.

Although we only marginally resolve the neutrinosphere with $\approx 1$ grid point per scale height at radii where $\tau_{\bar \nu_e} \sim 1$, the main effect of this deficiency is on the properties (luminosity, mean energy) of the escaping neutrino flux (Paper I). Although the asymptotic electron fraction of the wind is very sensitive to these properties, the main focus of this study is on the effects of a strong magnetic field for an otherwise fixed neutrino radiation field (and the neutrino properties do not depend strongly on the magnetic field; see Tab.~\ref{tab:models}).  Furthermore, at larger radii, specifically in the gain region where net neutrino heating launches the wind, we do sufficiently resolve the optical depth scale height.  For the purposes of this study, the resolution of our simulations is therefore sufficient to capture magnetic field effects and bulk wind dynamics.
 
\subsection{Hemisphere Symmetry Assumption}
\label{appendix:symmetry}

\begin{figure}
\centering
    \includegraphics[width=0.48\textwidth]{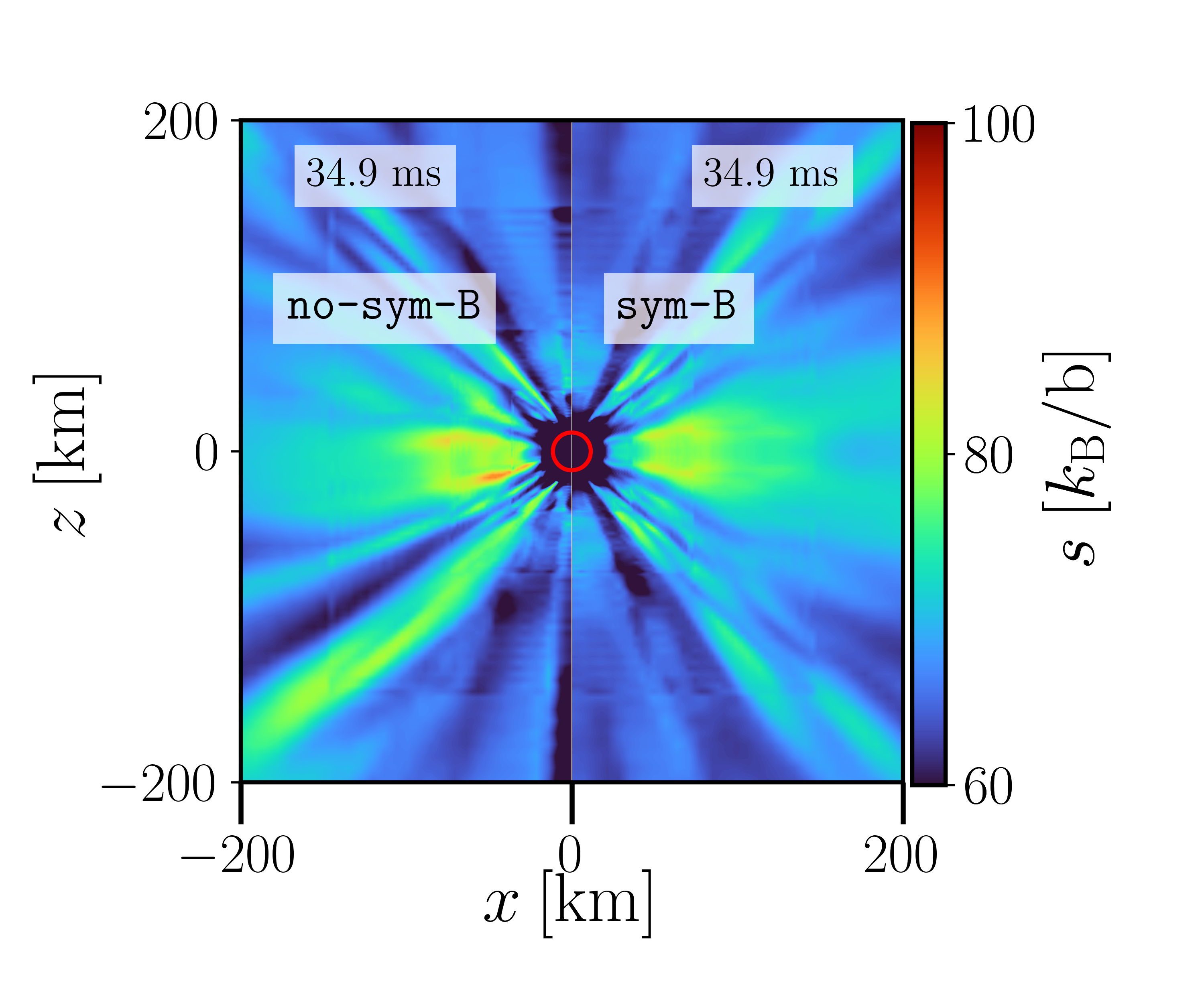}
    \includegraphics[width=0.48\textwidth]{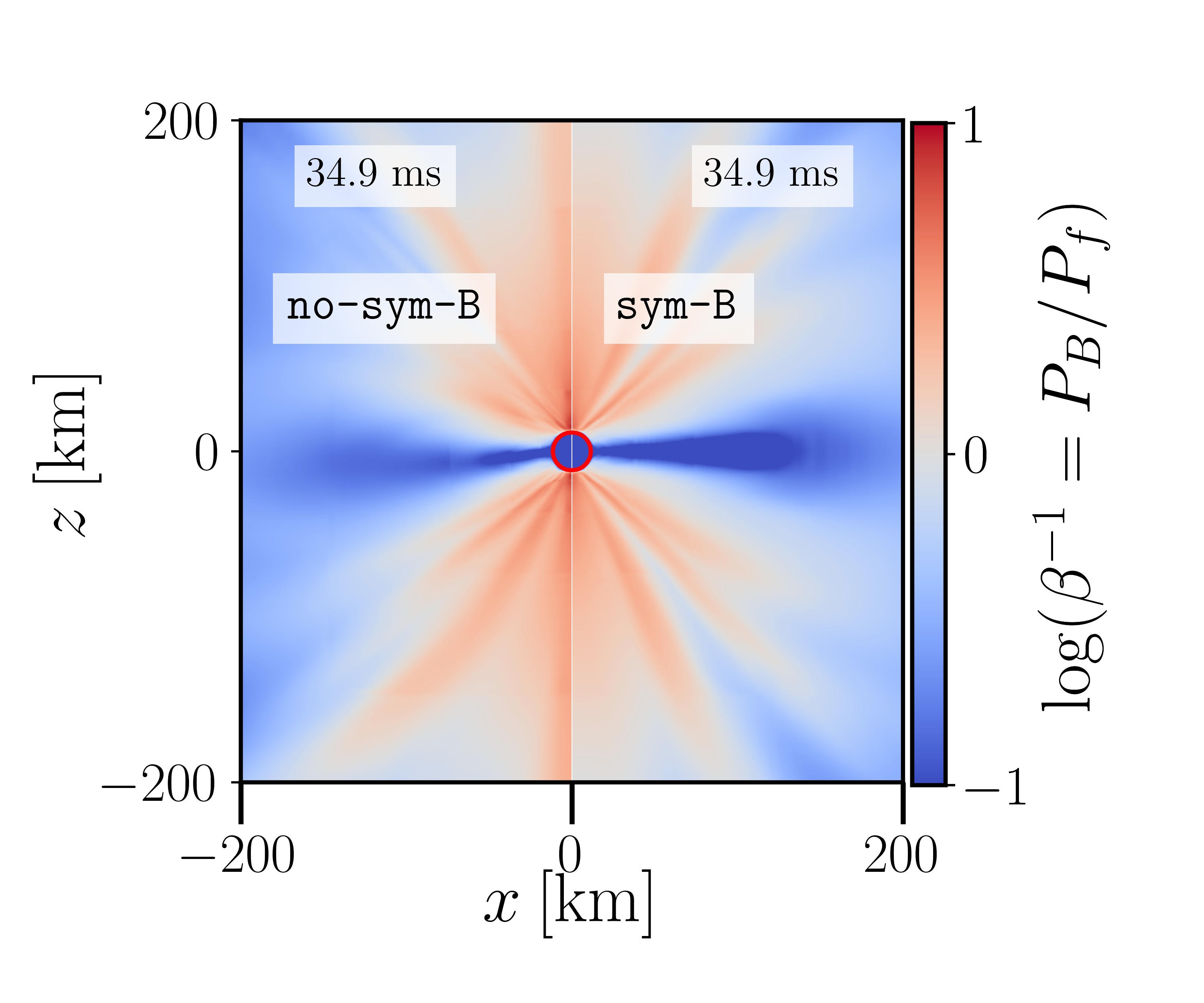}
    \caption{Comparison of wind properties with and without imposing reflection symmetry across the equatorial plane.  Shown are wind cross-sections in the plane of the magnetic dipole axis for entropy (left panel) and magnetic-to-fluid pressure ratio (right panel) time-averaged over the interval $t\simeq30- 35$ ms after the B-field is initialized. For each panel: the simulation is run across the full $\theta \in [0,\pi]$ domain (left; model \texttt{no-sym-B}); reflection symmetry across the $z=0$ plane is employed (right; model \texttt{sym-B}).}
\label{fig:ref_sym}
\end{figure}

We perform two otherwise identical simulations, with and without imposing reflection symmetry across the equatorial ($z=0$) plane, to check that the results of the half-hemisphere simulations presented in this paper are independent of the use of this assumption (\texttt{no-sym-B} and \texttt{sym-B}, Tab.~\ref{tab:models}). The two simulations use the same refinement level box sizes as those of our fiducial models, but with the resolution of the smallest refinement level being 450 m rather than 150 m, for reasons of computational expense associated with the full-domain simulations.  The set-up of the two models is similar to the fiducial magnetized models: after reaching roughly steady-state wind properties with zero magnetic field, we initialize the dipole magnetic field of strength $B_{\rm S} \approx 2.2\E{15}$ G and further evolve the models for $\simeq 40$ ms.

Broadly, the temporal evolution of the two simulations are qualitatively similar to those of the \texttt{hi-B} model ($B_{\rm S} = 2.5\times 10^{15}$ G): a thin reconnection layer with low magnetic-to-fluid pressure ratio $\beta^{-1}$ forms in the equatorial plane, and high $\beta^{-1}$ in the polar region due to magnetic field lines being torn open and approaching a split monopole-like solution in that region.
The half and full-domain simulations are even more similar to each other. Figure~\ref{fig:ref_sym} shows a snapshot comparing the magnetic-to-fluid pressure ratio roughly 35 ms after the magnetic field has been initialized.  
Although the low $\beta^{-1}$ equatorial current sheet/reconnection layer becomes slightly warped in the full-domain simulation (perhaps due to reconnection related-instabilities), the reconnection layer appears nearly symmetric and similar to the half-domain simulation when time-averaged over a 20 ms interval.  We conclude that while our half-domain simulation may miss some features of the dynamics near the equatorial plane, the time-averaged wind properties will not be greatly effected by this simplification. 

\subsection{Toroidal Field}
\label{appendix:toroidal}

\begin{figure}
\centering
    \includegraphics[width=0.9\textwidth]{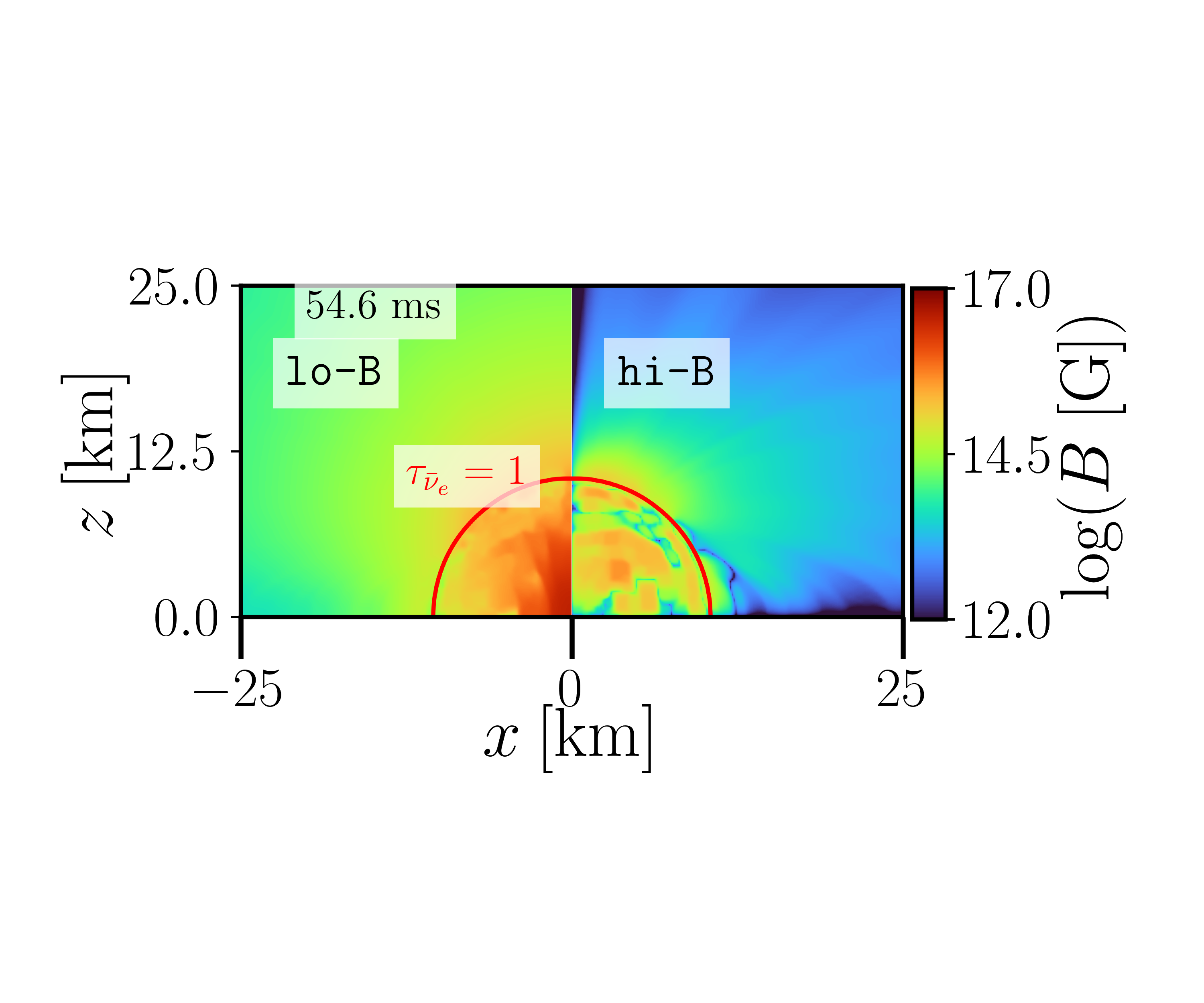}
    \caption{Strength of the poloidal (left) and toroidal (right) magnetic field components for model \texttt{hi-B} approximately 55\,ms after initialization of the magnetic field. The red contour indicates the location of the neutrinosphere.}
\label{fig:Btor}
\end{figure}

Absent any large-scale non-radial motions (due, e.g., to rotation), or the presence of other non-axisymmetric instabilities such as convection, we should expect the strength of the toroidal component of the magnetic field, $B_{\rm T}$, to remain highly subdominant compared to the poloidal field, $B_{\rm P}$.  We check this expectation in Fig.~\ref{fig:Btor} by showing $B_T$ and $B_{\rm P}$ from a snapshot of our \texttt{hi-B} model at 60 ms.  The ratio $B_{\rm T}/B_{\rm P}$ achieves a maximum value $10^{-1}$ inside the PNS, but has typical values $\lesssim 10^{-2}$ outside the neutrinosphere everywhere else on the grid.  As expected, the toroidal field should thus have no appreciable impact on the wind dynamics in the case of a non-rotating PNS.

\subsection{Landau Level Effects}
\label{appendix:landau}
\begin{figure}
\centering
    \includegraphics[width=0.49\textwidth]{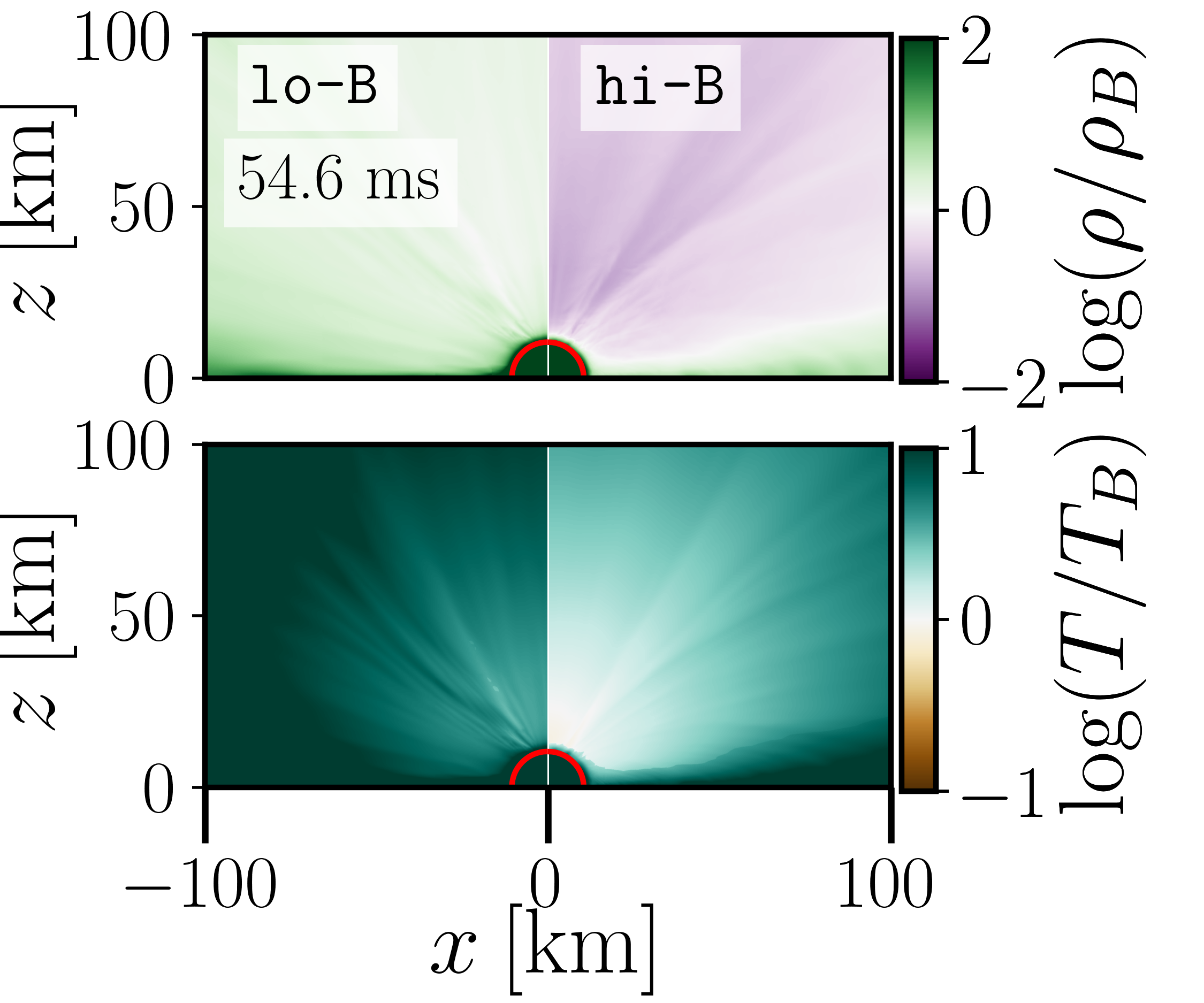}
    \caption{Cross-section of the $y=0$ magnetic dipole axis plane for the \texttt{lo-B} (left) and \texttt{hi-B} run (right) $\approx\!55$ ms after the magnetic field is initialized, showing the ratio of temperature $T$ (top) and density $\rho$ (bottom) to their respective critical values $T_B$ and $\rho_B$ below which quantizing effects of the magnetic field on the EOS and weak interactions are expected (Eqs.~\ref{eq:temp_B} and \ref{eq:rho_B}). Since $T/T_B > 1$ everywhere, electrons and positrons populate many Landau levels, even in the regime where $\rho < \rho_B$. The magnetic winds are thus in the non-quantizing regime, justifying our negligence of magnetic-field effects onto the EOS and weak interactions. The red contour indicates the $\tau_{\bar \nu_e}=1$ neutrinosphere surface.}
\label{fig:landau}
\end{figure}

Strong magnetic fields modify neutrino absorption and emission rates as well as the EOS via quantization of electron and positron energy levels resulting from the quantization into Landau levels of $e^\pm$ motion transverse to the magnetic field \citep{lai_neutrino_1998,duan_neutrino_2004,duan_rates_2005}. Such modifications become irrelevant for temperatures $T\gtrsim T_B$ or $\rho \gg \rho_B$, where $T_B$ is a critical temperature
\be
T_B =
    \begin{cases}
        \frac{m_e c^2}{k_B} \left( \sqrt{ \frac{2B}{B_{\rm Q}}+1}-1  \right) & \text{ for}\, \rho \le \rho_B\\
        \frac{\hbar \omega_c}{k_B} (1+x_F^2)^{-1/2} & \text{for}\, \rho \gg \rho_B
    \end{cases} \label{eq:temp_B}
\ee
and $\rho_B$ is a critical density
\be
\rho_B = 2.23 \times 10^9 \left(\frac{Y_e}{0.1} \right)^{-1} \left(\frac{B}{10^{15} ~\mathrm{G}} \right)^{3/2} \mathrm{g~cm}^{-3}, \label{eq:rho_B}
\ee
defined as the density below which only the ground Landau level is populated by electrons \citep{harding_physics_2006,haensel_neutron_2007}. Here, $m_e$ is the electron mass, $c$ is the speed of light, $\omega_c=eB/(m_e c)$ is the cyclotron frequency, $x_F = (\hbar/m_e c)(3\pi^2 Y_e \rho/m_{\rm u})^{1/3}$ is the normalized relativistic Fermi momentum, $m_{\rm u}$ the atomic mass unit, and $B_{\rm Q}=4.414 \times 10^{13}$ G is the critical QED magnetic field strength (obtained by equating the cyclotron energy of an electron to $m_ec^2$). If either the density or temperature is larger than their respective critical values, the $e^\pm$ distributions extend over many Landau levels and the magnetic field does not have a quantizing effect. Figure~\ref{fig:landau} shows the ratios $T/T_B$ and $\rho/\rho_B$ of our magnetized runs once a stationary PNS wind has emerged. Since $T>T_B$ everywhere, the magnetic field is non-quantizing, even in the polar regions where $\rho < \rho_B$, justifying our assumptions regarding the impact of magnetic fields on the EOS and weak interactions.

\subsection{Divergence-free Constraint Violations}
\label{appendix:divB}

\begin{figure}
\centering
    \includegraphics[width=0.99\textwidth]{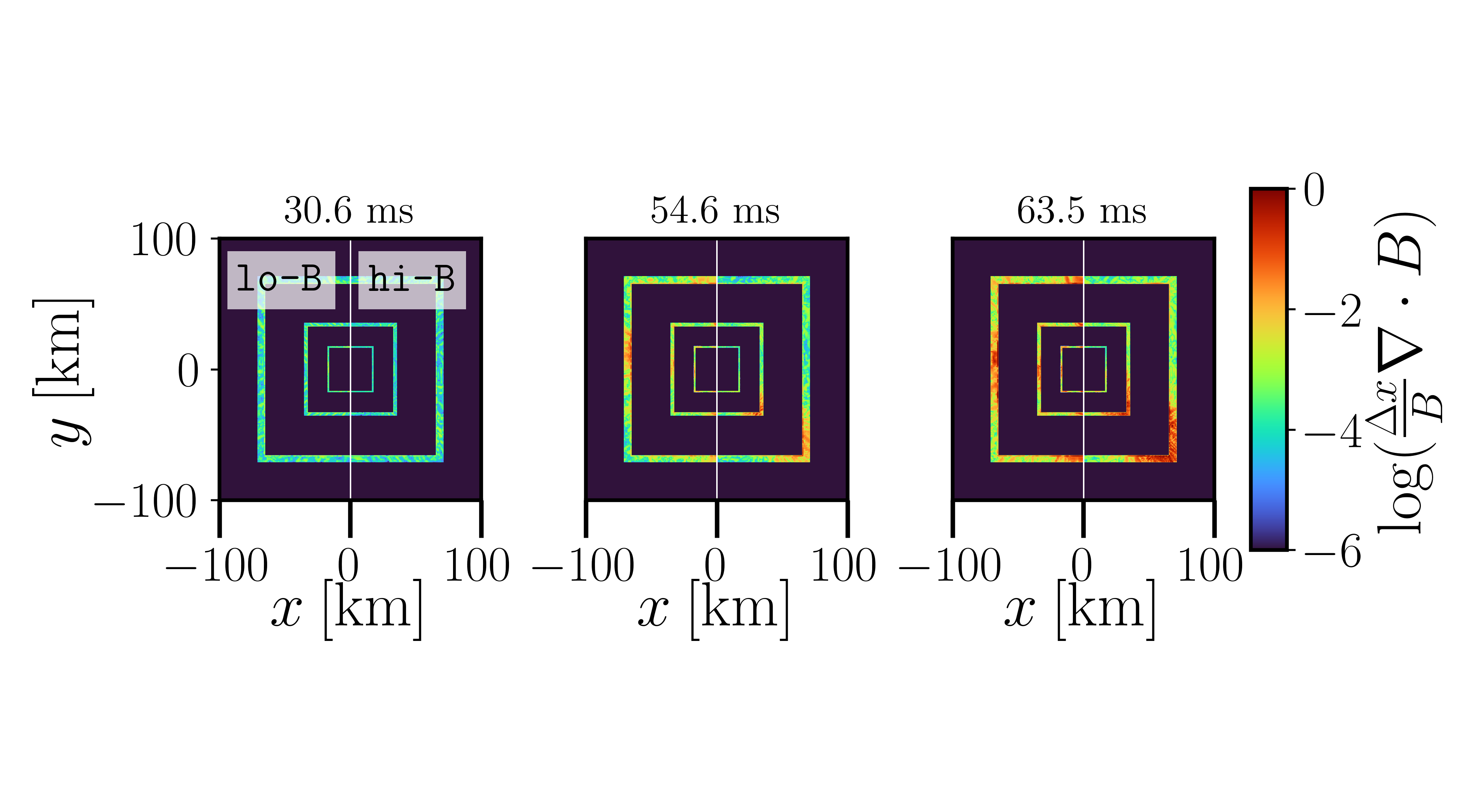}
    \vspace{-20mm}
    \caption{Equatorial plane ($z=0$) cross-section at 3 snapshots in time  ($\approx$ 31 ms, 55 ms, 64 ms after dipole magnetic field initialization) for models \texttt{lo-B} (shown in the $x<0$ domain) and \texttt{hi-B} (shown in the $x>0$ domain). The color represents the relative level of $\nabla \cdot \mathbf{B}$ violations, where $\Delta x$ is the grid spacing, and $B$ is the overall strength of the magnetic field. While the constraint transport scheme maintains $\nabla \cdot \mathbf{B}$ to machine precision interior to refinement level boundaries, spurious violations are introduced at refinement boundaries due to interpolation operations.
    %While violations are kept under control within the simulation domain, violations occur most prominently on the refinement level boundaries.
    }
\label{fig:divB}
\end{figure}

The magnetic field in our simulations is evolved using the FluxCT method \citep{toth_nablacdot_2000} to maintain the solenoidal constraint $\nabla \cdot \mathbf{B} = 0$ during evolution. While interior to refinement level boundaries this constrained transport algorithm preserves $\nabla \cdot \mathbf{B}$ to machine precision, spurious violations are introduced over time at refinement level boundaries due to interpolation during prolongation and restriction operations, which do not preserve $\nabla \cdot \mathbf{B}$ to machine precision. Although we use additional overlap zones at refinement boundaries to minimize the impact of spurious violations on the evolution of the system, significant violations in ghost zones may impact the solution quantitatively beyond a certain timescale, which depends on the exact grid setup and the physical system. In the current case, spurious violations introduced during the first $\sim150$\,ms of `pre-evolution' to establish a stationary, essentially non-magnetized wind (cf.~Sec.~\ref{sec:methods}), as well as violations introduced during magnetic field initialization and subsequent strongly magnetized evolution give rise to spurious accumulation of errors at refinement boundaries to the $\sim\,1\%$ level by $\gtrsim\!60$\,ms after the large-scale dipole magnetic field is initialized. Figure~\ref{fig:divB} illustrates the spurious growth of $\nabla \cdot \mathbf{B}$ at refinement boundaries as a function of time after initialization of the dipole field. At $\approx 64$\,ms and onward, we consider the accumulated errors at refinement level boundaries in the wind zone ($\sim\!60$\,km) prohibitive to fully trusting results from subsequent evolution and thus choose not to include subsequent simulation data into our analyses. In order to prevent effects of spurious $\nabla\cdot \mathbf{B}$ violations on the closed zone material while maintaining the resolution requirements at the neutrinosphere (Appendix \ref{appendix:resolution}) would require increasing the innermost box to $\approx\!100$\,km. The associated increase in computational cost by a factor of $\sim\!(100/15)^3\approx 300$ would render these simulations computationally infeasible. 

%We use the constrained transport formalism to prevent spurious growth of the $\nabla \cdot B =0$ constraint. Any errors accumulated only reach the $\sim 1 \%$ level by $t\approx 55$ ms. 

\end{document}